\documentclass[preprint,12pt]{elsarticle}

\usepackage[utf8]{inputenc}

\usepackage{amssymb}
\usepackage{amsmath}
\usepackage{amsthm}
\usepackage{eucal}
\usepackage{mathrsfs}
\usepackage{subfigure}
\usepackage{fix-cm}
\usepackage{bm}
\usepackage{amsfonts}
\usepackage{enumerate}
\usepackage{color}

\usepackage{ulem}
\usepackage{empheq}

\def\dO{\ \text{d}\hspace{-0.3mm} \Omega}

\def\x{\bm{x}}

\journal{}

\begin{document}

\begin{frontmatter}

%% Title, authors and addresses

%% use the tnoteref command within \title for footnotes;
%% use the tnotetext command for theassociated footnote;
%% use the fnref command within \author or \address for footnotes;
%% use the fntext command for theassociated footnote;
%% use the corref command within \author for corresponding author footnotes;
%% use the cortext command for theassociated footnote;
%% use the ead command for the email address,
%% and the form \ead[url] for the home page:
%% \title{Title\tnoteref{label1}}
%% \tnotetext[label1]{}
%% \author{Name\corref{cor1}\fnref{label2}}
%% \ead{email address}
%% \ead[url]{home page}
%% \fntext[label2]{}
%% \cortext[cor1]{}
%% \affiliation{organization={},
%%             addressline={},
%%             city={},
%%             postcode={},
%%             state={},
%%             country={}}
%% \fntext[label3]{}

\title{Optimal design of unimorph-type cantilevered piezoelectric energy harvesters using level set-based topology optimization by considering manufacturability
%Optimal design of piezoelectric energy harvesters using level set-based topology optimization considering manufacturability
}

%% use optional labels to link authors explicitly to addresses:
%% \author[label1,label2]{}
%% \affiliation[label1]{organization={},
%%             addressline={},
%%             city={},
%%             postcode={},
%%             state={},
%%             country={}}
%%
%% \affiliation[label2]{organization={},
%%             addressline={},
%%             city={},
%%             postcode={},
%%             state={},
%%             country={}}

\author[inst0,inst1]{Ken Miyajima}

\affiliation[inst0]{organization={Osaka Research Institute of Industrial Science and Technology},%Department and Organization
            addressline={Ayumino--2--7--1}, 
            city={Izumi--city},
            postcode={594-1157}, 
            state={Osaka},
            country={Japan}}

\affiliation[inst1]{organization={Department of Mechanical Engineering, Graduate School of Engineering, The University of Tokyo},%Department and Organization
            addressline={Yayoi 2--11--16}, 
            city={Bunkyo--ku},
            postcode={113--8656}, 
            state={Tokyo},
            country={Japan}}

\author[inst1,inst2]{Takayuki Yamada}

\affiliation[inst2]{organization={Department of Strategic Studies, Institute of Engineering Innovation, Graduate School of Engineering, The University of Tokyo},%Department and Organization
            addressline={Yayoi 2--11--16}, 
            city={Bunkyo--ku},
            postcode={113--8656}, 
            state={Tokyo},
            country={Japan}}

\begin{abstract}
%% Text of abstract
In this study, we propose a design methodology for a piezoelectric energy-harvesting device optimized for maximal power generation at a designated frequency using topology optimization.
The proposed methodology is adapted to the design of a unimorph-type piezoelectric energy harvester, wherein a piezoelectric film is affixed to a singular side of a silicon cantilever beam.
Both the substrate and the piezoelectric film components undergo concurrent optimization.
Constraints are imposed to ensure that the resultant design is amenable to microfabrication, with specific emphasis on 
the 
etchability of piezoelectric energy harvesters.
Several numerical examples are provided to validate the efficacy of the proposed method.
The results show that the proposed method yields optimized substrate and piezoelectric designs with an enhanced electromechanical coupling coefficient, while allowing the eigenfrequency of the device and the minimum output voltage to be set to the desired values.
Furthermore, the proposed method can provide solutions that satisfy the cross-sectional shape, substrate-dependent, and minimum output voltage constraints.
The solutions obtained by the proposed method are manufacturable in the field of microfabrication.
\end{abstract}

%%Graphical abstract
%\begin{graphicalabstract}
%\includegraphics{grabs}
%\end{graphicalabstract}

%%Research highlights
%\begin{highlights}
%\item Research highlight 1
%\item Research highlight 2
%\end{highlights}


\end{frontmatter}

%% \linenumbers

%% main text

\section{Introduction}
\label{sec:intro}

%% For citations use: 
%%       \citet{<label>} ==> Jones et al. [21]
%%       \cite{<label>} ==> [21]
%%

Piezoelectric energy harvesters convert mechanical energy into electrical energy through the piezoelectric effect.
These devices have gained significant attention in recent years owing to their potential to scavenge energy from various sources such as vibrations \cite{jeon2005mems,roundy2003study} and fluid flow \cite{yun2016vibrating}.
The design of a piezoelectric energy harvester plays a critical role in determining its efficiency and performance \cite{erturk2011piezoelectric}.
Various piezoelectric energy harvester designs, including cantilever beams, multilayered structures, and {cylindrical} structures, have been proposed and studied \cite{cook2008powering,beeby2006energy}.
The unimorph cantilevered piezoelectric energy harvester has several advantages over other piezoelectric energy harvester designs.
A unimorph cantilevered piezoelectric energy harvester consists of a cantilever beam with one end fixed and the other end free to move. The beam is typically composed of two primary domains: a substrate domain and a piezoelectric material domain. Additionally, it includes electrode domains for electrical energy extraction.
When the device is subjected to external vibrations or mechanical loads, the beam bends, causing mechanical strain in the piezoelectric domain. This strain generates electrical energy through the piezoelectric effect.
The simple and compact design of this harvester makes it suitable for applications where size and weight are critical factors, such as in portable or wearable devices.
Furthermore, the unimorph cantilevered design allows easy integration with other systems, such as sensors or wireless transmitters \cite{saxena2021design}.
However, the performance of a unimorph cantilevered piezoelectric energy harvester is limited by the amount of mechanical energy harvested from external vibrations.
To overcome these limitations, researchers have explored the use of advanced materials and optimization techniques to improve their performance. 
Some studies have investigated piezoelectric materials with improved electromechanical coupling factors to enhance the piezoelectric effect and improve the efficiency of energy conversion \cite{yoshimura2013piezoelectric,murakami2013development}.
Other researchers have focused on designing the geometry of the cantilever beam and piezoelectric element for maximum energy output. 
For instance, rectangular, triangular, and trapezoidal shapes have been proposed for unimorph piezoelectric vibration energy harvesters \cite{glynne2001towards}.
However, as these structures have simple geometries, significant improvements can be achieved using topological optimization techniques \cite{BENDSOE1988197}.

Among the early forays in the topology optimization of piezoelectric materials, the study by Silva et al. \cite{NELLISILVA199849} stood out as a pioneering study. 
This study explored the optimal design of periodic piezocomposites and underscored the profound impact of the unit cell topology on the performance of piezoelectric materials. 
Their insights provided a foundational framework for subsequent advancements in the field.
The optimization of a piezoelectric energy harvester requires meticulous attention to numerous parameters, including the geometry and material properties of the piezoelectric element, the resonant frequency of the device, and the electrical load impedance.
Zeng et al. \cite{Zheng2009} introduced a topology optimization technique to maximize the harvesting performance of piezoelectric energy harvesters.
Rupp et al. \cite{rupp2009design} 
conceived a design for a piezoelectric energy harvester integrated with an external electrical circuit with due regard to the electrical load impedance. 
K$\mathrm{\ddot{o}}$gl et al. \cite{kogl2005topology} advanced the ``PEAMAP-P" analysis model, specifically tailored for the analytical modeling of piezoelectric materials. 
Their design paradigm singularly targeted the piezoelectric material domain and deliberately omitted the substrate domain. 
Kang et al. \cite{kang2008integrated} considered both the piezoelectric material and electrode domains in their design.
 Adding a unique perspective, Zhang et al. \cite{ZHANG2014200} focused on the dynamic responses in piezoelectric structures, emphasizing transient load optimization. 
Their approach highlights the need for a precisely tailored design in active-control smart structures.
Chen et al. \cite{CHEN20102532} and Luo et al. \cite{luo2009design} designed tubular energy harvesting devices made of multiple materials.
Notably, they employed level set-based topology optimization to eliminate grayscale discrepancies between the material and void domains.
He et al. \cite{he2022topology} and  de Almeida et al. \cite{de2019topology} focused on designing the thickness profile of a piezoelectric energy harvester by considering the piezoelectric material, substrate, and electrode domains. 
In addition, Kim et al. \cite{kim2014topology} optimized both the substrate and piezoelectric material domains.
They delineated distinct objective functions for each domain and sequentially optimized the substrate and piezoelectric film structures.
Similarly, in the field of acoustic energy focusing, Yoon et al. \cite{YOON2018600} demonstrated the application of topology optimization for piezoelectric acoustic focusers. 
By considering the intricate interactions between the electric, mechanical, and acoustic phenomena, their approach offers a more comprehensive understanding and optimization of piezoelectric devices in acoustically challenging environments.
In a notable advancement, Salas et al. \cite{SALAS2018223} extended topology optimization of piezoelectric actuators beyond traditional metal substrates and piezoelectric films to include the optimization of fibers and polarization in laminated piezocomposite multi-entry actuators (LAPAs). 
This holistic approach presents a significant challenge in the design of sophisticated actuators with enhanced performance capabilities.
To further develop their methods, Salas et al. \cite{SALAS2021114010} introduced the hybrid interpolation model for fiber orientation (HYIMFO) method, which harmonizes continuous and discrete fiber orientation optimization for the LAPA. 
This hybrid approach represents a progressive stride in the optimization of complex multi-material actuators.

However, these methodologies often yield designs that face inherent manufacturing challenges. Consequently, various strategies have been proposed to realize manufacturable designs through topology optimization.
Vatanabe et al. \cite{vatanabe2016topology}, Michailidis et al. \cite{michailidis2014manufacturing}, and Sato et al. \cite{sato2017manufacturability} developed advanced methodologies to assess manufacturability.
Vatanabe et al. \cite{vatanabe2016topology} proposed a unified projection-based approach to incorporate manufacturing constraints into topology optimization, enabling the generation of manufacturable designs with minimum member size, minimum hole size, symmetry, and other geometric constraints. This method ensures that the optimized structures are not only optimal in terms of performance but also feasible for manufacturing processes like casting, milling, turning, forging, and rolling.
Michailidis et al. \cite{michailidis2014manufacturing} focused on multi-phase shape and topology optimization via a level-set method, incorporating manufacturing constraints to ensure feasible designs for practical applications.
Sato et al. \cite{sato2017manufacturability} developed a method 
for evaluating manufacturability in molding processes using fictitious physical models to ensure that the geometrical features of molded parts are amenable to manufacturing. 
Building upon the advancements in manufacturability assessments, various studies have proposed constraints related to manufacturability specifically in the context of additive manufacturing.
Liu et al. \cite{liu2018current} reviewed the current trends and future directions in topology optimization for additive manufacturing, highlighting the need for support structure optimization and addressing issues related to overhangs and material anisotropy.
Yamada et al. \cite{yamada2022topology} focused on additive manufacturing and proposed a topology optimization method that addressed closed cavity exclusion constraints and enhanced the design process of additive manufacturing. 
Tajima et al. \cite{TAJIMA2023116415} introduced a coupled fictitious physics model aimed at improving the convergence in topology optimization by integrating the objectives of both the fictitious physical and mechanical models, particularly for additive manufacturing with geometric constraints.
These methodologies collectively underscore the importance of incorporating manufacturability considerations into topology optimization, which is essential for practical applications in various manufacturing processes.
The level set-based topology optimization championed by 
Yamada et al. \cite{YAMADA20102876} affords precision in modulating the geometrical intricacy of the resultant optimal configurations.
Moreover, earlier studies developed various methodologies for multi-material topology optimization, such as those by Yulin and Xiaoming \cite{yulin2004level} and Wang and Wang \cite{wang2004color}, which laid the groundwork for subsequent advancements in this field.
Building on this foundation, Noda et al. \cite{noda2022extended} augmented the level set method to address the specific challenges of multi-material topology optimization.
This innovation ensures a meticulous orchestration of the complexity intrinsic to each deployed material, culminating in a multi-material assembly that is  conducive to straightforward manufacturing.
However, the previously underscored methodologies are predominantly conceptualized for traditional, macroscopic machining techniques such as milling. 
Therefore, their direct application to piezoelectric energy harvesters, typically realized through microfabrication techniques such as etching \cite{murakami2013development,aramaki2019demonstration}, may not be feasible.

In this study, we introduce a methodology for the optimal design of piezoelectric energy harvesters by employing topology optimization based on the level set method. 
This approach emphasizes manufacturability in the field of microfabrication, with etching being the predominant process under consideration. 
The design criteria for piezoelectric energy harvesters are delineated based on the operational frequency and a threshold for the excitation voltage output. 
Furthermore, the proposed method uniquely facilitates the concurrent optimization of three-dimensional structures within both the substrate and piezoelectric material domains. 
The resulting designs not only satisfy the stipulated specifications but are also readily adaptable to microfabrication procedures.

The remainder of this paper is organized as follows:
In Section \ref{sec:form}, the architectural framework of the designated device
is described.
After this initial exposition, in that section we further elaborate on the optimal design approach, articulating the pivotal design criteria, including the eigenfrequency specifications and constraints regarding the minimum output voltage.
In Section \ref{sec:constraint}, we explain the formulation of constraints pertinent to manufacturability.
Specifically, we introduce two constraints to enhance manufacturability: ensuring a consistent cross-sectional shape across each domain and  stipulation that no piezoelectric material is positioned without an accompanying substrate.
In Section \ref{sec:impli}, the implementation of the proposed method is described.
In Section \ref{sec:example},  numerical illustrations employing a benchmark model to test the efficacy of the introduced method are detailed.
Finally, in Section \ref{sec:conclusion}, the conclusions of this study are summarized.
Additionally, the methodologies tailored for computing the output voltage are detailed in \ref{secA:solv}, and the sensitivity analyses are discussed in \ref{secA:sens}.

\section{Formulation for the optimal design of a piezoelectric energy harvester}
\label{sec:form}
%-----------------------------------------------------------
\subsection{Unimorph cantilevered piezoelectric energy harvester}
\label{subsec:uni}

\begin{figure*}[htb]
	\begin{center}
		\includegraphics[height=3cm]{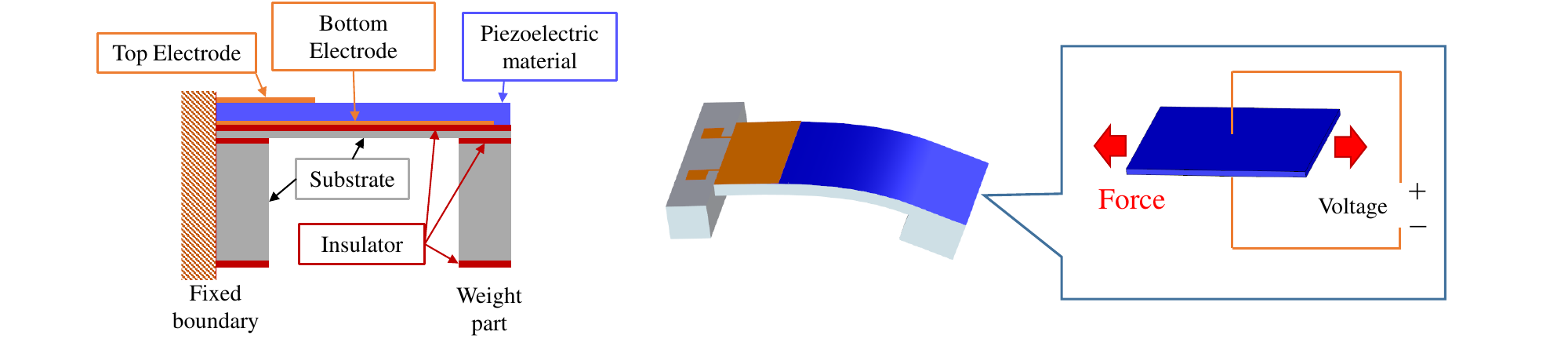}
		\caption{Concept of a piezoelectric energy harvester}
		\label{fig:concept_piezo}
	\end{center}
\end{figure*}

This study predominantly
 focuses 
 on the unimorph cantilevered energy harvester. 
A unimorph cantilevered piezoelectric energy harvester comprises three main components: substrate, piezoelectric material, and weight. 
These elements are depicted in Figure \ref{fig:concept_piezo}, which shows the theoretical structure of a unimorph cantilevered piezoelectric vibration energy harvester.
A unimorph cantilevered piezoelectric energy harvester consists of a cantilever beam with one end fixed and the other end free to move. 
The beam has a structure typically composed of three primary domains: a substrate domain, a piezoelectric material domain, and a weight domain at the free end. 
Additionally, it includes electrode domains for electrical energy extraction. 
The substrate domain is nonpiezoelectric and serves as the base layer to which the piezoelectric material is bonded. 
In this study, the substrate is composed of silicon, and the piezoelectric material utilized is lead zirconate titanate (PZT).
To ensure electrical insulation between the electrodes and the substrate domain, an insulating layer is formed on both the upper and lower surfaces of the substrate domain. 
When the device is subjected to external vibrations or mechanical loads, the weight causes the beam to bend more significantly, inducing mechanical strain in the piezoelectric domain. 
This strain generates electrical energy through the piezoelectric effect.
The presence of the weight enhances the bending of the beam, resulting in greater deformation of the piezoelectric layer and thus a higher electrical output.

The piezoelectric effect is a phenomenon where certain materials, such as quartz and PZT, 
generate an electric charge in response to applied mechanical stress. In a piezoelectric material, when mechanical strain is induced, an electric potential is developed across the material due to the displacement of charge carriers. 
This property allows piezoelectric materials to be used in energy harvesting devices, converting mechanical energy from vibrations into electrical energy.

The unique advantage of the unimorph design is its capability to produce a superior electrical output for a given mechanical input compared to a conventional cantilevered piezoelectric energy harvester. 
This superiority is attributed to the mechanical amplification effect induced by the substrate domain, which results in an enhanced deformation of the piezoelectric layer for a given input vibration.
From a design viewpoint, the unimorph-type harvester features a simpler structure with a single piezoelectric layer, allowing for easier fabrication and reduced manufacturing costs. 
The design flexibility is also increased as both the substrate and the piezoelectric layer can be independently etched, allowing for more complex geometries and optimized vibration characteristics.

%-----------------------------------
\subsection{Formulation of piezoelectric phenomena}
\label{subsec:pie}

The governing equations for a piezoelectric energy harvester are articulated in the following set of simultaneous piezoelectric equations, as elaborated in \cite{erturk2011piezoelectric}:

\begin{empheq}[left=\empheqlbrace]{align}
    \label{eq:gov_tensor_form}
    \begin{split}
        \sigma_{ij} &= C^E_{ijkl} s_{kl}  + e_{kij} \varphi_{,k} \\
        d_i &= e_{ikl} s_{kl}  - {\varepsilon}^S_{ik} \varphi_{,k},
    \end{split}
\end{empheq}

Here, $\sigma$ is the stress tensor; $d$ is the electric displacement tensor; $s$ represents the strain tensor; ${C}^E$ is the elasticity tensor at a constant electric field, and we assume that $C^E$ represents an isotropic linear elastic material; $e$ represents the piezoelectric stress constant tensor; $\varphi$ is the electric potential; 
$\varphi_{,k}$ is a derivative with respect to the $k$ th variable; 
and ${\varepsilon}^S$ is the dielectric constant tensor at a constant strain.

The piezoelectric coefficient tensor $e$ plays a crucial role in these equations, as it describes the coupling between the mechanical and electrical states of the material. 
In the 
second 
equation, $e_{kij}$ represents the piezoelectric effect, where mechanical strain $s_{kl}$ induces an electric displacement $d_i$. 
Conversely, in the 
first 
equation, the application of an electric field $\varphi_{,k}$ results in mechanical stress $\sigma_{ij}$. 
This bidirectional coupling is the hallmark of piezoelectric materials, allowing them to convert mechanical energy into electrical energy and vice versa. 
Therefore, the piezoelectric effect creates a strong interaction between the electric field and the elastic field within the material, enabling it to function efficiently as an energy harvester, sensor, or actuator.

\subsubsection{Matrix and vector formulation of piezoelectric phenomena}

To facilitate numerical computations and implementation, it is often convenient to convert these tensor equations into matrix and vector forms. 
This involves representing the strain tensor $s$ in its vector form, commonly using Voigt notation, and expressing the piezoelectric stress constant tensor $e$ and other relevant tensors in matrix forms.

Using Voigt notation, the strain tensor $s$ can be represented as a column vector $\bm{s}(\bm{u})$, and the elasticity tensor ${C}^E$ and piezoelectric stress constant tensor $e$ are transformed accordingly. 
The electric field $E_k$ is related to the electric potential $\varphi$ by $E_k = -\varphi_{,k}$. 
Consequently, the piezoelectric equations in matrix-vector forms are given by:

\begin{empheq}[left=\empheqlbrace]{align}
    \label{eq:gov_strong_form}
    \begin{split}
        \bm{\sigma} &= \bm{C}^E \bm{s}(\bm{u})  + \bm{e}^T \bm{\nabla} \varphi \\
        \bm{d} &= \bm{e} \bm{s}(\bm{u})  - \bm{\varepsilon}^S \bm{\nabla} \varphi.
    \end{split}
\end{empheq}
In these equations, $\bm{\sigma}$ represents the stress vector (6$\times$1); $\bm{d}$ is the electric displacement vector (3$\times$1); $\bm{s}(\bm{u})$ is the strain vector (6$\times$1); $\bm{C}^E$ is the elasticity matrix (6$\times$6); $\bm{e}$ is the piezoelectric coefficient matrix (3$\times$6); $\bm{\varepsilon}^S$ is the dielectric constant matrix (3$\times$3); and $\bm{\nabla} \varphi$ is the electric field vector (3$\times$1) derived from the electric potential $\varphi$.

In this study, we primarily
 focus 
 on the steady-state vibration of a piezoelectric energy harvester under harmonic excitation at a frequency $\omega$. 
Within this context, the displacement vector $\bm{u}$ and the electric potential $\varphi$ are designated as state variables. 

Assuming a time-harmonic form for the displacement and electric potential, such that $\bm{u}(t) = \bm{u} e^{i\omega t}$ and $\varphi(t) = \varphi e^{i\omega t}$, the time-dependent piezoelectric equations with external forcing can be expressed as follows:
\begin{empheq}[left=\empheqlbrace]{align}
        \label{eq:time_dependent_form}
        \begin{split}
            \bm{C}^E \bm{s}(\bm{u}) + \bm{e}^T \bm{\nabla} \varphi &= \rho \frac{\partial^2 \bm{u}}{\partial t^2} + \hat{\bm{f}} \\
            \bm{e} \bm{s}(\bm{u}) - \bm{\varepsilon}^S \bm{\nabla} \varphi &= \hat{\bm{q}}.
        \end{split}
\end{empheq}
Here, $\rho$ denotes the material's density, $\hat{\bm{f}}$ represents the external forcing function, and $\hat{\bm{q}}$ indicates the externally provided electric charge.

By substituting $\bm{u}(t) = \bm{u} e^{i\omega t}$ and $\varphi(t) = \varphi e^{i\omega t}$ into these equations, we obtain the following:
\begin{empheq}[left=\empheqlbrace]{align}
    \begin{split}
        \bm{C}^E \bm{s}(\bm{u}) - \omega^2 \rho \bm{u}  + \bm{e}^T\bm{\nabla}\varphi
        = \hat{\bm{f}}\\
        \bm{e}\bm{s}(\bm{u}) - \bm{\varepsilon}^S\bm{\nabla}\varphi = \hat{\bm{q}}.  \label{eq:gov_strong_form_eigen}
    \end{split}
\end{empheq}
Note that the term $-\omega^2 \rho \bm{u}$ arises from the second derivative of the harmonic displacement $\bm{u} e^{i\omega t}$ with respect to time, representing the inertia force.

To determine the eigenfrequencies of the system, we consider the free vibration case with no external forces or charges. This leads to the eigenvalue problem of the piezoelectric field under open-circuit and short-circuit conditions.
In the open-circuit condition, the top and bottom electrodes are not connected, implying the electrical resistance between the electrodes is assumed to be infinite. 
This configuration inhibits any charge transfer between the electrodes, thereby preserving the voltage difference induced by the polarization of the piezoelectric material. 
Conversely, in the short-circuit condition, the top and bottom electrodes are connected with zero electrical resistance. 
This allows any charge accumulation due to the polarization of the piezoelectric material to be immediately neutralized, resulting in no electric field within the piezoelectric material. 
As indicated by the equations (\ref{eq:gov_strong_form_eigen}), the change in the electric field within the piezoelectric material directly affects its stiffness. 
In practical applications, the device typically operates with a finite electrical resistance connected to electronic circuits, resulting in a state between these two conditions.

Under open-circuit conditions (where $\hat{\bm{q}}=0$), 
the eigenvalue problem is expressed as follows:

\begin{empheq}[left=\empheqlbrace]{align}
    \begin{split}
    	\bm{C}^E \bm{s}(\bm{u}_{oci}) - \omega_{oci}^2 \rho \bm{u}_{oci}    	+\bm{e}^T\bm{\nabla}\varphi_{oci} 
    	&=0\\
    	\bm{e}\bm{s}(\bm{u}_{oci})
    	-  \bm{\varepsilon}^S\bm{\nabla}\varphi_{oci} &=0.
        \label{eq:strong_open}
    \end{split}
\end{empheq}
Here, $\omega_{oci}$ is the $i$-th mode eigenfrequency under open-circuit conditions, $\bm{u}_{oci}$ is the eigenvector corresponding to the eigenvalue $\omega_{oci}$, and $\varphi_{oci}$ denotes the electric potential for the $i$-th vibration mode under open-circuit conditions. 
Under open-circuit conditions, the piezoelectric effect increases the stiffness, leading to a higher eigenfrequency $\omega_{oci}$.
Conversely, under short-circuit conditions (where $\bm{\nabla}\varphi = 0$), 
the eigenvalue problem
 reduces to the following equation
 , unaffected by the influence of the piezoelectric material.
\begin{align}
	 \bm{C}^E \bm{s}(\bm{u}_{sci}) - \omega_{sci}^2 \rho  \bm{u}_{sci}  
	= 0.
    \label{eq:strong_short}
\end{align}
Here, $\omega_{sci}$ is the $i$-th mode eigenfrequency under short-circuit conditions, and $\bm{u}_{sci}$ is the eigenvector corresponding to the eigenvalue $\omega_{sci}$.
Under short-circuit conditions, 
the stiffness is reduced due to the neutralization of the piezoelectric effect, resulting in a lower eigenfrequency $\omega_{sci}$.

Comparing the eigenfrequencies under these two conditions provides insight into the impact of the piezoelectric effect on the device's behavior. 
The difference in stiffness between open-circuit and short-circuit conditions results in a difference between the eigenfrequencies $\omega_{oci}$ and $\omega_{sci}$. 
By comparing these eigenfrequencies, the electromechanical coupling coefficient $k^2_i$ can be calculated, serving as a measure of the effectiveness of the piezoelectric material in converting mechanical energy into electrical energy. 
The electromechanical coupling coefficient for each vibration mode is given by the following equation \cite{hagood1991damping}:
\begin{align}
    k^2_i = \frac{\omega_{oci}^2 - \omega_{sci}^2}{\omega_{oci}^2}.
    \label{eq:f_ome}
\end{align}

In the context of a unimorph cantilevered piezoelectric energy harvester design, the eigenfrequency of the device is established as a crucial design parameter
for the following reasons. First, these devices generate maximum output when they are resonating at their eigenfrequency, as the energy conversion efficiency is highest in this state. 
Second, by tuning the eigenfrequency to match the frequency of the ambient vibration source, the harvester can achieve optimal performance at its installation location by effectively capturing and converting the environmental vibration energy into electrical energy \cite{erturk2011piezoelectric}.
Consequently, in this study, we considered strategies for concurrently aligning the eigenfrequency with the desired value.
Adopting the objective function proposed by 
Yamada et al. \cite{Yamada2012}, we normalized the eigenfrequency specification problem with respect to the $i$-th mode desired frequency $\bar{\omega}_i$ and formulated the objective function $F_{\omega}$ as follows:
\begin{align}
    \label{eq:obj_ev}
    F_{\omega}=&\sum_{i=1}^n \frac{| \omega_{oci} - \bar{\omega}_i |^2 }{\bar{\omega}_i^2},
\end{align}
where $n$ donates the number of modes to evaluation. 
In this study, for the design of a 3-dimensional beam, we set $n=4$. 
This is because we need to consider not only the first mode of vibration but also the second mode and bending vibrations in the direction perpendicular to the primary vibration direction, as well as torsional vibrations of the beam, which collectively require evaluating up to the fourth mode.

In this study, we aim to design a high-performance piezoelectric energy harvester by maximizing the electromechanical coupling coefficient $k^2_i$. 
To transform the maximization problem into a minimization problem, we formulated the objective function $F_k$ 
 using the electromechanical coupling coefficient $k_i$ as follows:
\begin{align}
    \label{eq:obj_piezo}
    F_k = \sum_{i=1}^n \frac{1}{k^2_i \bar{\omega}_i^2} =\sum_{i=1}^n \frac{\omega_{oci}^2}{(\omega_{oci}^2 - \omega_{sci}^2)\bar{\omega}_i^2} .
\end{align}
By weighting the performance $k_i$ by the modes $\bar{\omega}_i$, we emphasized the lower-order modes.

For a unimorph cantilevered piezoelectric energy harvester design, we perform the simultaneous optimization of the two objective functions $F_k$ and $F_\omega$ using a weighting factor $\alpha$ as follows:
\begin{align}
\underset{\Omega}{\text{inf}} \quad F = \alpha F_k + (1-\alpha) F_\omega.
\end{align}
This approach allows us to obtain an optimized design that achieves enhanced power generation efficiency among solutions with the desired eigenfrequency. 
Furthermore, to optimize both the substrate and piezoelectric material regions, we employ the following method.

We formulate an optimization problem that simultaneously minimizes distinct objective functions {for different material domains of the energy harvester, specifically, $F_{sb}$ for the substrate material domain $\Omega_{sb}$ and $F_{pe}$ for the piezoelectric material domain $\Omega_{pe}$. 
The substrate material domain $\Omega_{sb}$ consists of silicon, while the piezoelectric material domain $\Omega_{pe}$ is composed of
 PZT. 
The eigenvectors $\bm{u}_{oci}$ and $\bm{u}_{sci}$, as well as the electric potential $\varphi_{oci}$, are continuous across the boundary between $\Omega_{sb}$ and $\Omega_{pe}$.

The governing equations for each domain under open-circuit and short-circuit conditions are as follows. 
In the piezoelectric material domain $\Omega_{pe}$, the governing equations under open-circuit conditions are:
\begin{align}
    \bm{C}^E_{pe} \bm{s}(\bm{u}_{oci}) - \omega_{oci}^2 \rho_{pe} \bm{u}_{oci} + \bm{e}^T \bm{\nabla} \varphi_{oci} = 0 \quad \text{in} \; \Omega_{pe} \\
    \bm{e} \bm{s}(\bm{u}_{oci}) - \bm{\varepsilon}^S \bm{\nabla} \varphi_{oci} = 0 \quad \text{in} \; \Omega_{pe}
\end{align}
and under short-circuit conditions 
it 
is:
\begin{align}
    \bm{C}^E_{pe} \bm{s}(\bm{u}_{sci}) - \omega_{sci}^2 \rho_{pe} \bm{u}_{sci} = 0 \quad \text{in} \; \Omega_{pe}.
\end{align}
Similarly, in the substrate material domain $\Omega_{sb}$, the governing equations under open-circuit conditions is:
\begin{align}
    \bm{C}^E_{sb} \bm{s}(\bm{u}_{oci}) - \omega_{oci}^2 \rho_{sb} \bm{u}_{oci} = 0 \quad \text{in} \; \Omega_{sb},
\end{align}
and under short-circuit conditions is:
\begin{align}
    \bm{C}^E_{sb} \bm{s}(\bm{u}_{sci}) - \omega_{sci}^2 \rho_{sb} \bm{u}_{sci} = 0 \quad \text{in} \; \Omega_{sb}.
\end{align}
Here, $\bm{C}^E_{pe}$ and $\bm{C}^E_{sb}$ denote the elastic matrices of the piezoelectric and substrate materials, respectively, while $\rho_{pe}$ and $\rho_{sb}$ represent the densities of the piezoelectric and substrate materials. 
The eigenvector fields $\bm{u}_{oci}$ and $\bm{u}_{sci}$ are shared between the piezoelectric and substrate domains under open-circuit and short-circuit conditions, respectively.

This optimization incorporates weighting factors $\alpha_{sb}$ and $\alpha_{pe}$ to balance the contributions of each domain to the overall objective. The formulation of the objective functions $F_{sb}$ and $F_{pe}$, and the incorporation of the weighting factors $\alpha_{sb}$ and $\alpha_{pe}$, are demonstrated below, following the methodology proposed by} Kim and Lee \cite{kim2014topology}:
\begin{align}
    \label{eq:form_piezo_ev_gen}
    \begin{array}{rll}
        \underset{\Omega_{sb}}{\text{inf}} \quad & F_{sb} = \alpha_{sb} F_k + (1-\alpha_{sb}) F_{\omega} &\\
        \underset{\Omega_{pe}}{\text{inf}} \quad & F_{pe} = \alpha_{pe} F_k + (1-\alpha_{pe}) F_{\omega}& \\
        \text{subject to} \quad &{ \bm{C}^E_{pe} \bm{s}(\bm{u}_{oci}) - \omega_{oci}^2 \rho_{pe} \bm{u}_{oci} + \bm{e}^T \bm{\nabla} \varphi_{oci} = 0 }&\quad{ \text{in} \; \Omega_{pe}} \\
        & {\bm{e} \bm{s}(\bm{u}_{oci}) - \bm{\varepsilon}^S \bm{\nabla} \varphi_{oci} = 0} &\quad {\text{in} \; \Omega_{pe}} \\
        & {\bm{C}^E_{pe} \bm{s}(\bm{u}_{sci}) - \omega_{sci}^2 \rho_{pe} \bm{u}_{sci} = 0} &\quad {\text{in} \; \Omega_{pe}} \\
        & {\bm{C}^E_{sb} \bm{s}(\bm{u}_{oci}) - \omega_{oci}^2 \rho_{sb} \bm{u}_{oci} = 0} &\quad {\text{in} \; \Omega_{sb}} \\
        & {\bm{C}^E_{sb} \bm{s}(\bm{u}_{sci}) - \omega_{sci}^2 \rho_{sb} \bm{u}_{sci} = 0} &\quad {\text{in} \; \Omega_{sb}}
    \end{array}
\end{align}

%-----------------------------------
\subsubsection{Minimum output voltage constraint}
\label{subsec:vconst}

Concomitant with the eigenfrequency, the output voltage is a pivotal performance metric in the design of piezoelectric energy harvesters. 
This attribute is significant because devices harnessing the generated energy often require a minimum viable output voltage. 
Owing to the physical constitution of the piezoelectric energy harvester, wherein the piezoelectric material is encapsulated between electrodes, the resultant output voltage from the vibratory motion is closely correlated not only with the charge induced on the electrodes but also with the capacitance between the piezoelectric material and the electrodes. 
A larger piezoelectric material area results in a greater charge, albeit at the cost of increased capacitance. 
Hence, the capacitance must be meticulously regulated to match the required output voltage.
Macroscopically, the output voltage $V_E$ is expressed as follows:
\begin{align}
    V_E = \frac{Q}{C_p} = \frac{\int_{\Omega_{pe}}\bm{n}_z \cdot \nabla \varphi_{n} \dO}{\int_{\Omega_{pe}} \dO/(\varepsilon_{z}  L_z^2)} ,
    \label{eq:def_voltage}
\end{align}
where $C_p$ signifies the capacitance of the piezoelectric energy harvester, $Q$ denotes the electrical charge elicited by the piezoelectric material, $\bm{n}_z$ represents a unit vector in the polarization direction, $\varepsilon_z$ is the 
%polarization direction component of the piezoelectric material's dielectric constant tensor
{component of the piezoelectric material's dielectric constant tensor in the polarization direction}, and $L_z$ is the thickness of the piezoelectric domain. 
{$\varphi_n$ denotes the electric potential calculated using the modal superposition method. Details of the modal superposition method are described in \ref{secA:solv}.}

The constraint on the minimum output voltage is formulated as follows:
\begin{align}
    V_E = \frac{\int_{\Omega_{pe}}\bm{n}_z \cdot \nabla \varphi_n \dO}{\int_{\Omega_{pe}} \dO/(\varepsilon_{z}  L_z^2)} \ge \bar{V}_{minV},\\
    \int_{\Omega_{pe}}\dO \le \frac{\varepsilon_{z}  L_z^2 \int_{\Omega_{pe}}\bm{n}_z \cdot \nabla \varphi_n \dO}{\bar{V}_{minV}},
    \label{eq:output_voltage_const}
\end{align}
where $\bar{V}_{minV}$ denotes the minimum output voltage.
This formulation allows the minimum output voltage constraint to be treated similarly to the volume constraint.

%-----------------------------------------------------------
\subsection{Concept of topology optimization}
\label{subsec:top}

Topology optimization is a type of structural optimization method {\cite{allaire2001shape,bendsoe2003topology,azegami2020shape}}.
Structural optimization is used to obtain a structure $\Omega$ that minimizes or maximizes an objective function.
The objective function often includes physical properties, such as stiffness \cite{BENDSOE1988197,allaire2001shape}, as well as thermal \cite{ZHOU201924, yamada2011level, JING201561, wu2019multi, miki2021topology}, fluid  \cite{Feppon2020topology, guan2023topology}, electromagnetic \cite{choi2009simultaneous, yamada2013topology}, and acoustic properties \cite{WADBRO2006420, sigmund2003systematic, noguchi2022topology, noguchi2021topology}.

{
\subsubsection{Formulation of topology optimization}
Typically in topology optimization methods}, the optimal structure is obtained under the assumption that the objective function satisfies the governing equations that describe the physical phenomena.
The governing equations are considered as constraints in the optimization problem, and the basic structural optimization problem can be formulated as follows:
\begin{align}
	\begin{split}
	   \underset{\Omega}{\text{inf}} \qquad &F(u,\Omega) = \int_{\Omega}f(u)\mathrm{d\Omega}\\
	   \text{subject to} \qquad &\text{the governing equations of the physical domain},\\
            &\text{additional constraint equations},
	   \label{strct_optim}
	\end{split}
\end{align}
where $u$ is the state variable obtained as the solution of the governing equation and $f(u)$ is the objective function.

Next, we 
consider 
the application of topology optimization to the structural optimization problem (\ref{strct_optim}).
We
 introduce 
 a domain $D\subset\mathbb{R}^n (n = 2 \; \text{or} \; 3)$ where the structure can be placed.
Here, the domain $D$ is called a fixed design domain because it does not change during the optimization process.
The fixed design domain is a domain filled with the structure (hereafter referred to as the material domain) and a domain not filled with the structure (hereafter referred to as the void domain), and these domains are expressed by the characteristic function $\chi$, which is defined as follows:
\begin{align}
	\chi(\bm{x}) := \left\{ 
	\begin{array}{ll}
    	1 \qquad \mathrm{for} \quad\bm{x}\in \Omega \\
    	0 \qquad \mathrm{for} \quad\bm{x}\in D\backslash\Omega , \\
	\end{array} 
	\right.
\end{align}
where the boundary between the material and void domains is included in the material domain.
Using the characteristic function $\chi$, the topology optimization problem can be formulated as follows:
\begin{align}
	\begin{split}
		\underset{\chi}{\text{inf}} \qquad &F(u,\chi) = \int_{D }f(u)\chi\mathrm{d\Omega}\\
		\text{subject to} \qquad &\text{the governing equation system},\\
            &\text{constraint equations} .
	   \label{top_optim}
	\end{split}
\end{align}
In topology optimization, the problem described in (\ref{strct_optim}) is reformulated as a material distribution problem. 
This allows for topological changes, such as an increase or decrease in the number of holes, during the optimization procedure.

However, topology optimization problems are commonly ill-posed \cite{allaire2001shape}; therefore, the space of admissible design should incorporate relaxation or regularization techniques to render the problem well-posed.
A typical method based on the relaxation in space of the admissible designs is the homogenization method \cite{BENDSOE1988197}.
In this study, we
 use 
 a level set-based topology optimization method \cite{YAMADA20102876} to transform an ill-posed problem into a well-posed problem.
In this method, the boundary surface of the material domain is represented by the isosurface of a scalar function called the level set function, and changes in the level set function represent changes in the shape of the material domain.
The topology optimization problem is regularized by ensuring the proper smoothness of the level set function.
This method is described in subsection \ref{subsec:lev}.

%-----------------------------------
\subsubsection{Level set-based topology optimization}
\label{subsec:lev}

In the level set-based method, the scalar function 
$\phi(\bm{x})\in H^1(D)$, 
also called the level set function, illustrated in the following equation is introduced to represent the shape {\cite{wang2003level}}:
\begin{eqnarray}
	\left\{ 
	\begin{array}{lll}
    	-1 \leq \phi(\bm{x}) <0 \qquad &\mathrm{for} \quad \bm{x} \in D\backslash\Omega \\
    	\phi(\bm{x})=0\qquad &\mathrm{for} \quad \bm{x} \in \partial\Omega \\
    	1 \geq \phi(\bm{x}) >0 \qquad &\mathrm{for} \quad \bm{x} \in \Omega, \\
	\end{array} 
	\right.
\end{eqnarray}
where $\partial \Omega$ denotes the boundaries between the material and void domains.
We redefined the characteristic function using the level set function as follows:
\begin{eqnarray}
	\chi_{\phi} (\phi(\bm{x})) := \left\{ 
	\begin{array}{ll}
    	1 \qquad \mathrm{for} \quad\phi(\bm{x}) \geq 0 \\
    	0 \qquad \mathrm{for} \quad\phi(\bm{x})< 0.  \\
	\end{array} 
	\right . 
	\label{eq:chi}
\end{eqnarray}
In the level set-based method, the topology optimization problem is formulated using the characteristic function defined in (\ref{eq:chi}).
However, as previously discussed, optimization that employs characteristic functions as design variables constitutes an ill-posed problem. 
Consequently, the level set-based method offers regularization to this optimization challenge, which is formulated as follows {\cite{allaire2004structural}}:
\begin{align}
	\begin{split}
		\underset{\phi}{\text{inf}} \qquad &F_R (u,\phi) = \int_{D}f(u)\chi_{\phi} (\phi)\mathrm{d\Omega} +  \int_{D}  \frac{1}{2} \tau|\nabla \phi|^2 \dO\\
	   \text{subject to} \qquad &\text{the governing equations of the physical domain},\\
            &\text{additional constraint equations},
	   \label{eq:levset_form}
	\end{split}
\end{align}
where $\tau \in \mathbb{R}_+$ is the  regularization coefficient.

The optimization problem represented by Eq. (\ref{eq:levset_form}), is reformulated using Lagrange’s method of undetermined multipliers. Let the Lagrangian be $L$ and the Lagrange multiplier of the governing equations be $v$, and 
that of the 
constraint equations be $\lambda$. The optimization problem 
is transformed into an 
unconstrained problem as 
follows: 
\begin{align}
    \underset{\phi}{\text{inf}}\quad L = \int_{D}f(u)\chi_{\phi} (\phi)\mathrm{d\Omega} +  \int_{D}  \frac{1}{2} \tau|\nabla \phi|^2 \dO + v \; J(u) + \lambda \; G(u). 
\end{align}

Next, we describe a method for updating the level set function.
Assuming that the level set function is a function of fictitious time $t$, it is updated by the reaction-diffusion equation as follows {\cite{YAMADA20102876}}:
\begin{align}
    \begin{split}
        \frac{\partial \phi(\bm{x},t)}{\partial t} &= -K(\phi)\left \{-\tilde{c} F' - \tau \nabla^2 \phi(\bm{x},t)  \right \} ,\\
        \tilde{c} :&= \frac{c \int_{D}\dO }{\int_{D} | F' | \dO } \;\; ,\\
        {F':} &{= \frac{\text{d} L}{\text{d} \phi}\;\; ,}
    \end{split}
    \label{eq:RDE}
\end{align}
where $K(\phi)\in \mathbb{R}_+$ is the proportionality coefficient, $c\in \mathbb{R}_+$ is the normalization coefficient, 
$L$ 
is the Lagrangian function, and $F'$ is the design sensitivity defined as gradient of the Lagrangian function.
$K(\phi)$ behaves as a coefficient to adjust the magnitude of sensitivity.
In this study, we set $K=1.0$ and $c=2.0$.

%-----------------------------------------------------------------------------------------------------------------------
\section{Formulation of constraints for manufacturability}
\label{sec:constraint}

\begin{figure}[htpb]
    \centering
    \includegraphics[width=\linewidth]{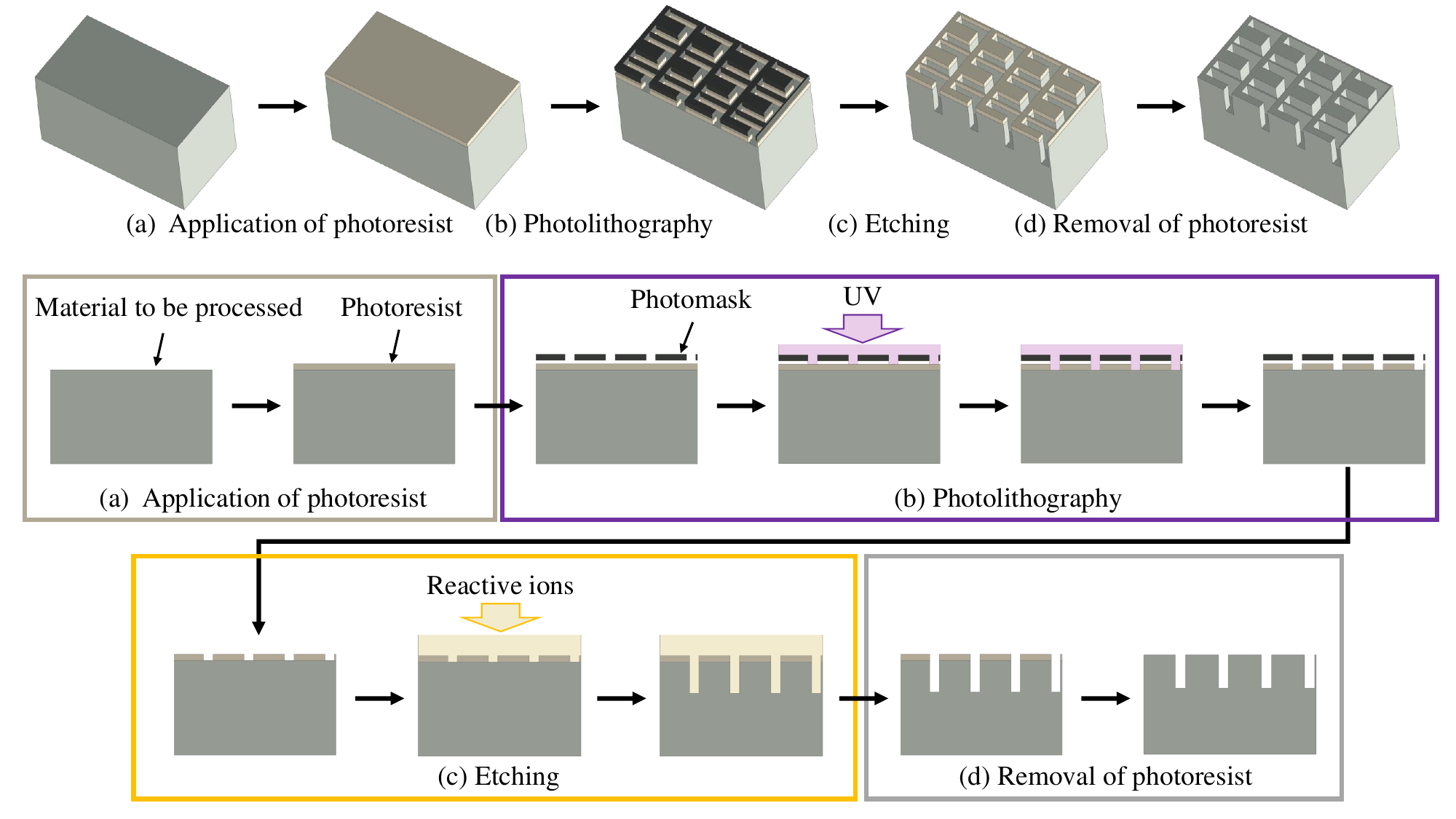} 
    \caption{Steps of the DRIE process creating precise microstructures.
        \textbf{(a) Application of resist:} A layer of photoresist is applied to the surface of the material.
        \textbf{(b) Photolithography:} 
        (1) A photomask is aligned over the photoresist-coated material. 
        (2) UV exposure through the photomask transfers the pattern onto the photoresist. 
        (3) The photoresist is developed, revealing the pattern.
        \textbf{(c) Etching:} 
        (1) Etching begins in the exposed areas by exposing the surface to reactive ions. 
        (2) Etching progresses to achieve the desired depth and pattern. 
        (3) Etching is completed with precise control in the vertical direction.
        \textbf{(d) Removal of resist:} The remaining photoresist is removed, leaving behind the etched pattern.}   
    \label{fig:etching_process}
\end{figure}

In this section, we describe the formulation of prerequisites related to manufacturing processes.
This study sets forth specific manufacturing criteria for piezoelectric energy harvesters, emphasizing configurations amenable to etching methodologies, predominantly utilizing the deep reactive ion etching (DRIE) apparatus \cite{murakami2013development, laermer1996method}.
Etching is a crucial step in the fabrication of microstructures, allowing for the precise removal of material. 
There are various etching techniques, primarily categorized into wet etching and dry etching. Wet etching involves chemical solutions to remove material, while dry etching uses plasma or gases. For the purposes of this study, we utilize the deep reactive ion etching (DRIE) method.
Figure \ref{fig:etching_process} outlines the DRIE process. DRIE is a highly anisotropic etching process that allows for the creation of complex structures in the planar direction using photolithography. 
This anisotropy means that etching can be precisely controlled in one direction, typically perpendicular to the substrate, while controlling etching in other directions is significantly more challenging. 
As a result, ensuring structural uniformity in the thickness direction is critical when designing structures using DRIE.

To this end, constraints are introduced to ensure structural uniformity in the thickness direction for both the piezoelectric and substrate domains.
Such constraints prevent the emergence of configurations with apertures oriented in oblique or orthogonal directions relative to the processing trajectory.

When etching is performed from only one side, the manufacturing constraints can be satisfied by imposing a constraint that the entire device has a uniform cross-section using a single level set function. 
This approach is mainly used for comparison in Section \ref{subsec:num_ex_1}.
On the other hand, when etching is performed from both sides, the structure does not need to have a uniform cross-section collectively. 
It is permissible for the piezoelectric material and the substrate to each have uniform cross-sections individually while maintaining overall manufacturability. 
By using the extended level set method \cite{noda2022extended}, a type of multi-material topology optimization method, we can satisfy the constraints for etching from both sides while ensuring design flexibility.

\begin{figure}[htpb]
    \centering
    \includegraphics[width=\linewidth]{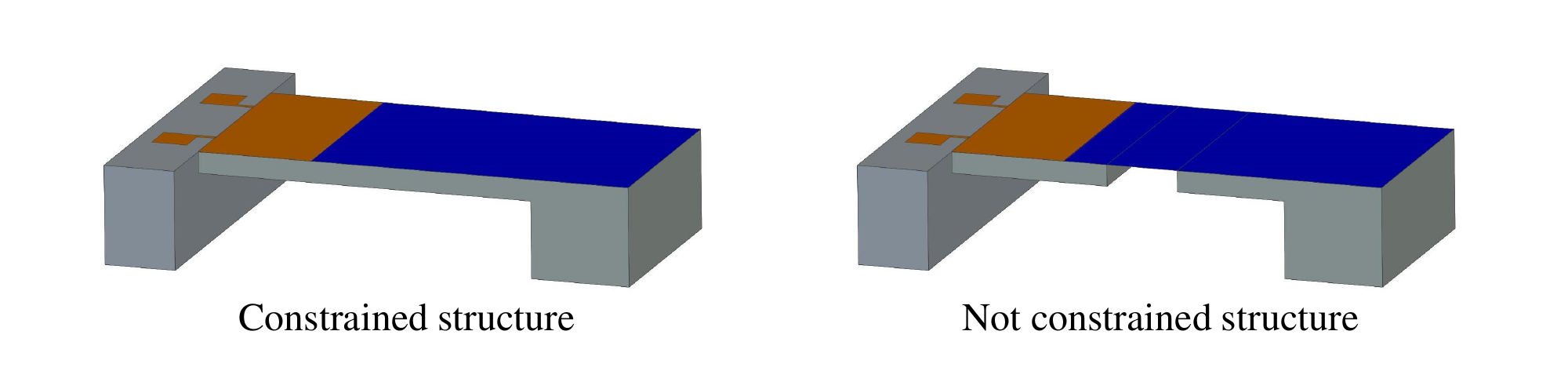} 
    \caption{Illustration of the substrate-dependent constraint on the piezoelectric domain. 
    The left side shows a "Constrained structure," where the piezoelectric material is properly supported by the substrate, ensuring mechanical stability and integrity.
    The right side shows a "Not constrained structure," where piezoelectric material is deposited without substrate support, leading to potential delamination or cracking. }   
    \label{fig:substrate_constraint}
\end{figure}

In addition to ensuring structural uniformity, it is crucial to impose a substrate-dependent constraint on the piezoelectric domain. 
This is to preclude the occurrence of piezoelectric materials in areas devoid of a substrate, which can result in mechanical instability and reduced device performance. 
Without this constraint, the piezoelectric material may be deposited in areas where it lacks adequate support from the substrate, leading to potential delamination or cracking during the fabrication process. 
Figure \ref{fig:substrate_constraint} illustrates the difference between constrained and not constrained structures. 
Ensuring that the piezoelectric materials are only present where there is a supporting substrate helps maintain the mechanical integrity and reliability of the device.

In this section, we present different methods to model and address the etching fabrication constraint. 
Section \ref{subsec:cross-sec} introduces methods to ensure that the cross-sectional shape of the device is uniform. 
Specifically, Section \ref{subsec:tau} describes the formulation of constraints for etching from a single side using the single level set function method, which is mainly used for comparative purposes. 
Section \ref{subsec:ex-ls} presents the formulation of constraints for etching from both sides using the extended level set function method. 
Section \ref{subsec:p} details the formulation of substrate-dependent constraints. 
Finally, Section \ref{subsec:form_opt} formulates the optimization problem incorporating both substrate-dependent constraints and the extended level set method.

%-----------------------------------------------------------
\subsection{Constraint: the same cross-sectional shape}
\label{subsec:cross-sec}

\begin{figure}[htpb]
        \centering
        \includegraphics[width=\linewidth]{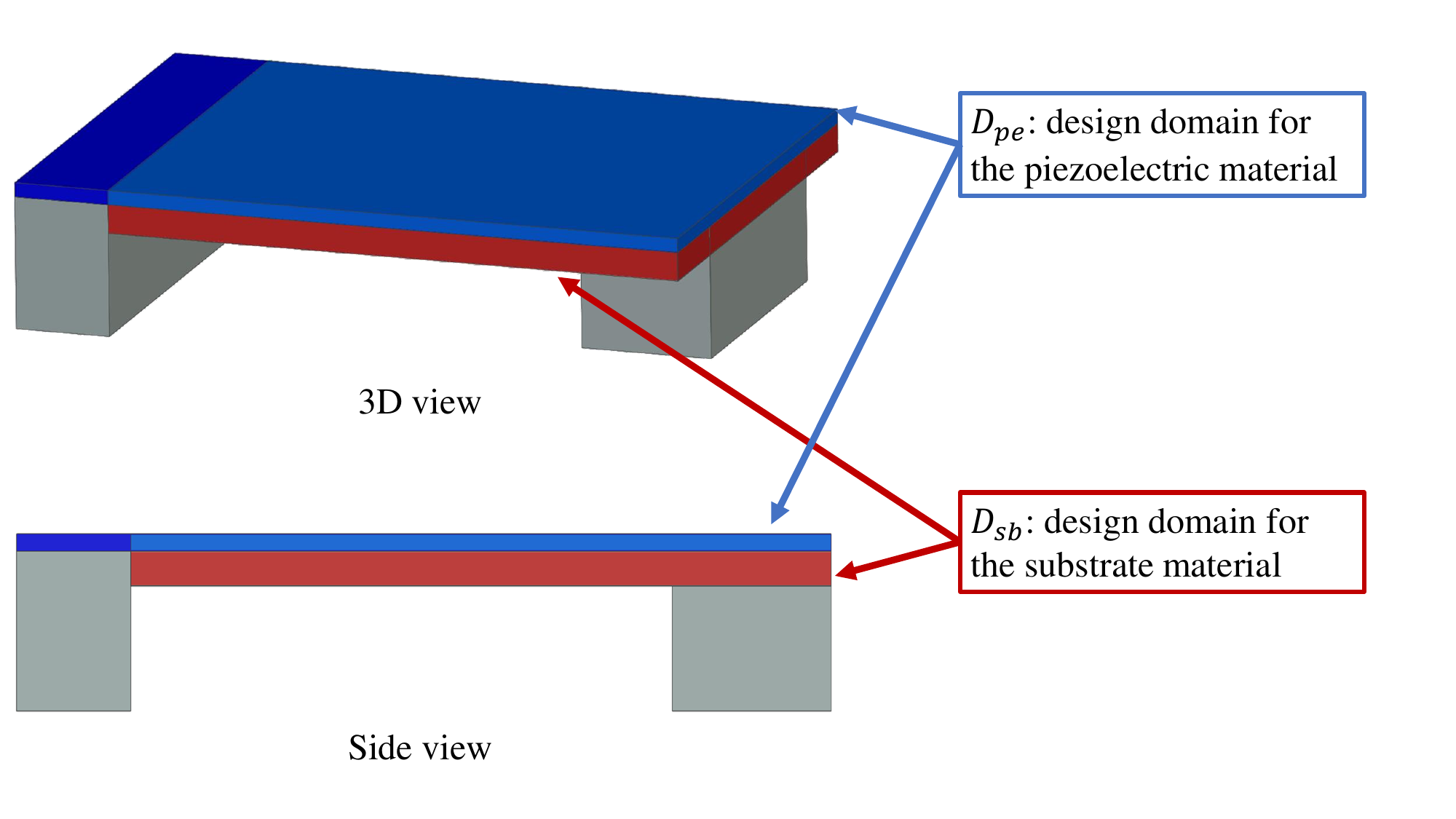} 
        \caption{
            The design domains used in this study. 
            \textbf{3D view:} Visualization of the 3D structure showing the division into the piezoelectric design domain ($D_{pe}$) and the substrate design domain ($D_{sb}$).
            \textbf{Side view:} Cross-sectional side view illustrating the components contained within each domain. 
            $D_{pe}$ is the design domain for the piezoelectric material, while $D_{sb}$ is design domain for the substrate material.
        }
        \label{fig:FDD}
    \end{figure}

\begin{figure}[htpb]
        \centering
        \includegraphics[width=\linewidth]{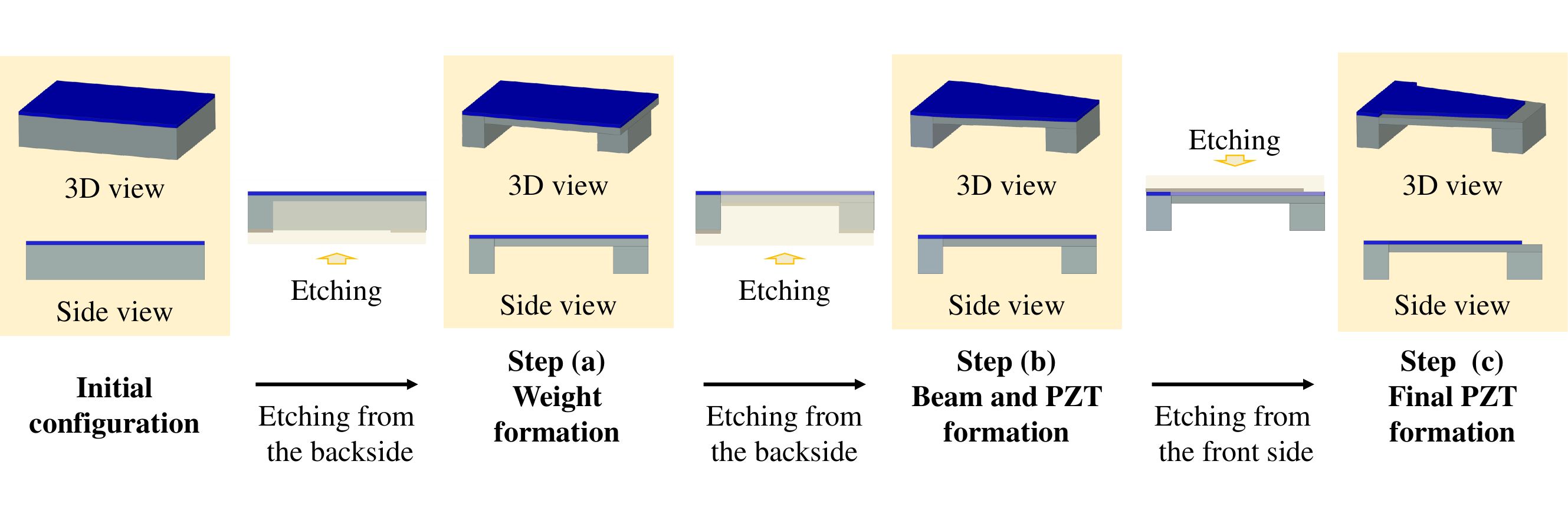} 
    \caption{
        Schematic illustration of the etching process for a unimorph cantilevered piezoelectric energy harvester. 
        \textbf{Initial configuration:} Before any etching process begins. 
        \textbf{Step (a) Weight formation:} The structure after etching from the backside to form the weight. 
        \textbf{Step (b) Beam and PZT formation:} The structure after etching from the backside to form the silicon beam and PZT shapes. This step represents single-side etching.
        \textbf{Step (c) Final PZT formation:} The structure after etching from the front side to form the PZT shape differing from the beam. This step represents double-side etching.
    }
        \label{fig:etching_example}
    \end{figure}

To ensure manufacturability via etching, it is crucial to impose constraints that maintain structural uniformity in the thickness direction. 
This section describes methods to achieve a uniform cross-sectional shape for the piezoelectric and substrate domains. 
The concept involves increasing the value of the regularization coefficient $\tau$ in the thickness direction to achieve uniform cross-sections. 
In this study, as shown in Figure \ref{fig:FDD}, we have two different design domains: $D_{sb}$ for the substrate material and $D_{pe}$ for the piezoelectric material. 
The entire design domain is defined as $D = D_{pe} \cup D_{sb}$. 
We address two approaches to enforce this constraint: one using a single level set function for etching from a single side, and the other using the extended level set method for etching from both sides. 
Figure \ref{fig:etching_example} illustrates the conceptual structures that satisfy each constraint. 
Structures as shown in Step (b) can be achieved by single-side etching, while structures as shown in Step (c) can be obtained through double-side etching.

%-----------------------------------------------------------
\subsubsection{Constraint: the same cross-sectional shape throughout the design domain}
\label{subsec:tau}

Within the framework of topology optimization obtained via the level set method, qualitative control over complexity can be regulated by the regularization coefficient $\tau$, as defined in Eq. (\ref{eq:levset_form}) {\cite{YAMADA20102876, allaire2004structural}}. 
To leverage this control over complexity, we adopt the approach used by Yamada et al. \cite{YAMADA20102876} to assign regularization coefficients that curtail the complexity in the thickness direction of the structure, thereby facilitating manufacturability via etching. 
In this study, we extend the regularization coefficient to a tensor $\bm{\tau}$ and
 modify 
 Eqs. (\ref{eq:levset_form}) and (\ref{eq:RDE}) as follows:

\begin{align}
	\begin{split}
    {
		\underset{\phi}{\text{inf}}}
        \qquad & {F_R (u,\phi) = \int_{D}f(u)\chi_{\phi} (\phi)\mathrm{d\Omega} +  \int_{D}  \frac{1}{2} |(\nabla\phi)^T \bm{\tau} \nabla\phi| \dO}\\
	   \text{subject to} \qquad &\text{the governing equations of the physical domain},\\
            &\text{additional constraint equations,}
	   \label{eq:levset_form_const}
	\end{split}\\
    \begin{split}
        \frac{\partial \phi(\bm{x},t)}{\partial t} = -K(\phi) \left\{ -\tilde{c} F' - {\nabla}^T \bm{\tau} {\nabla} \phi(\bm{x},t)  \right \} ,\\
        %\tilde{c} := \frac{C \int_{D}\mathrm{d\Omega} }{\int_{D} | d_t F | \mathrm{d\Omega} } ,
    \end{split}
    \label{eq:RDE2}
\end{align}
where
\begin{align}
    \bm{\tau} = \left[
    \begin{array}{lll}
        \tau_x & 0 & 0\\
        0 & \tau_y & 0\\
        0 & 0 & \tau_z
    \end{array}
    \right].
\end{align}
In this context, augmenting the component of the regularization coefficient tensor associated with the gradient in a specific direction of the level set function simplifies the complexity along that direction.
By defining the thickness direction as $z$ and assigning $\tau_x, \tau_y \ll \tau_z$, we can obtain a structure exhibiting minimal complexity in the thickness direction, thus enhancing manufacturability via etching.

To maintain a consistent cross-sectional shape throughout the design domain, the optimization problem is formulated using a single level set function $\phi$ and a single weight factor $\alpha$. In this study, since the physical properties of the object domain differ between the design domain of the piezoelectric material $D_{pe}$ and the design domain of the substrate $D_{sb}$, a constant function $\phi_{ps}$ is used to define the governing equations as follows:

\begin{eqnarray}
    \left\{
    \begin{array}{ll}
        \phi_{ps}(\bm{x})=1 &\mathrm{if}~~\bm{x}\in D_{pe}\\
        \phi_{ps}(\bm{x})=-1 &\mathrm{if}~~\bm{x}\in D_{sb},
        \label{eq:profile of chi_ps}
    \end{array}
    \right. 
\end{eqnarray}

\begin{align}
    \label{eq:form_piezo_single}
    \begin{array}{rll}
        \underset{\phi}{\text{inf}} \quad & F = \alpha F_k + (1-\alpha) F_{\omega} ,& \\
        \text{subject to} \quad &
        { 
            (\bm{C}^E_{pe}\chi_{\phi}(\phi_{ps}) + \bm{C}^E_{sb}\chi_{\phi}(1-\phi_{ps}) )\chi_{\phi}(\phi) \bm{s}(\bm{u}_{oci})
        } &\\
        &{\quad
            - \omega_{oci}^2 (\rho_{pe}\chi_{\phi}(\phi_{ps}) + \rho_{sb}(1-\chi_{\phi}(\phi_{ps}) ))\chi_{\phi}(\phi) \bm{u}_{oci} }&\\
        &{\quad    
            + \bm{e}^T \chi_{\phi}(\phi_{ps}) \bm{\nabla} \varphi_{oci}  
            = 0 
        }&\quad{ \text{in} \; D,} \\[0.5em]
        & {
            \bm{e}\chi_{\phi}(\phi_{ps})\chi_{\phi}(\phi) \bm{s}(\bm{u}_{oci}) - \bm{\varepsilon}^S\chi_{\phi}(\phi) \bm{\nabla} \varphi_{oci} 
            = 0
        } &\quad {\text{in} \; D,} \\[0.5em]
        & {
            (\bm{C}^E_{pe}\chi_{\phi}(\phi_{ps}) + \bm{C}^E_{sb}\chi_{\phi}(1-\phi_{ps}) )\chi_{\phi}(\phi) \bm{s}(\bm{u}_{sci}) 
        }&\\
        &{\quad
            - \omega_{sci}^2 (\rho_{pe}\chi_{\phi}(\phi_{ps}) + \rho_{sb}(1-\chi_{\phi}(\phi_{ps}) ))\chi_{\phi}(\phi) \bm{u}_{sci} = 0
        } &\quad {\text{in} \; D.}
    \end{array}
\end{align}

\subsubsection{Constraint: the same cross-sectional shape in each domain}
\label{subsec:ex-ls}

Although the method described above controls the complexity of the structure throughout the design domain, the substrate and piezoelectric components can be readily processed via etching, even if their structures differ. 
It is desirable to control the complexity in the thickness direction within the substrate and piezoelectric domains without affecting the structural differences at the interface of each domain.
To this end, in this study, we defined separate level set functions for the substrate and piezoelectric structures, invoking the extended level set method \cite{noda2022extended}, which is capable of handling multiple materials and updating each level set function concurrently. 
In the extended level set method, when dealing with a void domain and two types of object domains, making three types of domains in total, the level set function matrix is defined as follows:
\begin{align}
    \bm{\phi} = \left[
    \begin{array}{lll}
        \phi_{00} & \phi_{01} & \phi_{02}\\
        \phi_{10} & \phi_{11} & \phi_{12}\\
        \phi_{20} & \phi_{21} & \phi_{22}
    \end{array}
    \right],
\end{align}
\begin{align}
    \chi_{a}  &= \prod_{b\in[0, 1, 2]\backslash a} \chi_{\phi}(\phi_{ba}),\\
    \phi_{aa} &= 0 \qquad (a\in [0,1,2])\\
    \phi_{ab} &= -\phi_{ba} \qquad (a \ne b)
\end{align}

In this framework, $\phi_{p\psi}$ is defined as $\phi_{10}$, representing the boundary between the piezoelectric and void domains, and $\phi_{s\psi}$ is defined as $\phi_{20}$, representing the boundary between the substrate and void domains. Therefore, the level set function matrix can be redefined as follows:
\begin{align}
    \bm{\phi} &= \left[
    \begin{array}{lll}
        \phi_{\psi \psi} & \phi_{\psi p} & \phi_{\psi s}\\
        \phi_{p\psi} & \phi_{pp} & \phi_{ps}\\
        \phi_{s\psi} & \phi_{sp} & \phi_{ss}
    \end{array}
    \right],\\
    \phi_{\psi p} &= - \phi_{p \psi},\\
    \phi_{\psi s} &= - \phi_{s_\psi},\\
    \phi_{sp} &= -\phi_{ps},\\
    \phi_{\psi \psi} &= \phi_{pp} = \phi_{ss} = 0,
\end{align}
\begin{align}
    \chi_{p}(\bm{x}) &= \left\{
    \begin{array}{ll}
         1 &\mathrm{if}~~\bm{x}\in \Omega_{pe}\\
         0 & \mathrm{if}~~\bm{x}\in D\setminus {\Omega}_{pe} ,
        \label{eq:profile of chi_p}
    \end{array}
    \right. \\
    \chi_{s}(\bm{x}) &= 
    \left\{
    \begin{array}{ll}
         1 &\mathrm{if}~~\bm{x}\in \Omega_{sb}\\
         0 &\mathrm{if}~~\bm{x}\in D \setminus {\Omega}_{sb},\label{eq:profile of chi_s}
    \end{array}
    \right. \\
    \chi_{\psi}(\bm{x}) &= 
    \left\{
    \begin{array}{ll}
         1 &\mathrm{if}~~\bm{x}\in D\setminus(\Omega_{pe} \cup \Omega_{sb})\\
         0 &\mathrm{if}~~\bm{x}\in \Omega_{pe} \cup \Omega_{sb},
         \label{eq:profile of chi_psi}
    \end{array}
    \right. 
\end{align}
where $\chi_p$ denotes the piezoelectric material domain, $\chi_s$ denotes the substrate material domain, and $\chi_\psi$ denotes void domain.
The fixed design domains for the substrate and piezoelectric material, denoted as $D_{sb}$ and $D_{pe}$, respectively, are assumed to remain invariant throughout the optimization procedure.
The entire fixed design domain, denoted as $D$, encompasses both $D_{pe}$ and $D_{sb}$.
The level set function $\phi_{ps}$, representing the boundary between the domains $\omega_{pe}$ and $\omega_{sb}$, is defined as a constant value during the optimization process, as shown in the following equation:
\begin{eqnarray}
    \left\{
    \begin{array}{ll}
        \phi_{ps}(\bm{x})=1 &\mathrm{if}~~\bm{x}\in D_{pe}\\
        \phi_{ps}(\bm{x})=-1 &\mathrm{if}~~\bm{x}\in D_{sb},
        \label{eq:profile of chi_ps}
    \end{array}
    \right. 
\end{eqnarray}
The level set functions $\phi_{p\psi}$ and $\phi_{s\psi}$ are updated as follows:
\begin{align}
    \frac{\partial \phi_{p\psi}}{\partial t} = -K \left(-\tilde{c} F'_{p\psi} - \nabla (\bm{\tau}_{p\psi} \nabla \phi_{p\psi}) \right),\label{eq:reaction-diffusion eq_p}\\
    \frac{\partial \phi_{s\psi}}{\partial t} = -K \left(-\tilde{c} F'_{s\psi} - \nabla (\bm{\tau}_{s\psi} \nabla \phi_{s\psi}) \right),\label{eq:reaction-diffusion eq_s}
\end{align}
where $F'_{p\psi}$ and $F'_{s\psi}$ are the gradients of the objective function with respect to $\phi_{p\psi}$ and $\phi_{s\psi}$, respectively, and $\bm{\tau}_{p\psi}$ and $\bm{\tau}_{s\psi}$ are the regularization coefficient tensors for $\phi_{p\psi}$ and $\phi_{s\psi}$, respectively.
By individually defining the regularization coefficient tensors for the piezoelectric and substrate domains, the complexity in the thickness direction for each domain can be controlled. 
This approach allows for an increase in design flexibility without compromising manufacturability, compared to using a single level set function for the entire fixed design domain $D$.

%-----------------------------------------------------------
\subsection{Constraint: no piezoelectric material is placed without a substrate}
\label{subsec:p}

Herein, we elucidate an approach that allows the imposition of geometric constraints by using a fictitious physical function field denoted by ${\xi}$. 
The concept of fictitious physical models described by a steady-state anisotropic advection-diffusion equation is initially introduced by 
Sato et al. \cite{sato2017manufacturability}. 
A further elaboration of the fictitious physical model using the anisotropic diffusion equation is provided by 
Yamada et al. \cite{yamada2022topology}, 
which is described below.

\begin{align}
    \begin{split}
    {
        -\text{div} (\bm{\kappa}_{{\xi}} \nabla {\xi})  = 0 }&{\qquad \text{in} \quad D,}\\
    {
        {\xi}=\bar{{\xi}}_s}&{\qquad \text{in} \quad \Omega_{sb},
    }\\
        {\xi}=-\bar{{\xi}}_0  &\qquad \text{on}\quad \Gamma_{\xi},\\
        \nabla {\xi} \cdot \bm{n} = 0&\qquad \text{on}\quad\partial D \backslash \Gamma_{\xi},
        \label{eq:form_fict}
    \end{split}
\end{align}

\begin{align}
\bm{\kappa}_{\xi} = \left[
    \begin{array}{lll}
        \kappa_x & 0 & 0\\
        0 & \kappa_y & 0\\
        0 & 0 & \kappa_z
    \end{array}
    \right],
\end{align}
where $\partial D$ denotes the outer boundary of the entire fixed design domain $D$; $\bm{\kappa}_{\xi}$ represents the conductivity tensor of ${\xi}$; $\kappa_x$, $\kappa_y$, and $\kappa_z$ represent the anisotropic conductivities along the $x$, $y$, and $z$ directions, respectively; and $\bar{{\xi}}_s \in \mathbb{R}_+$ and $\bar{{\xi}}_0 \in \mathbb{R}_+$ are the constants for the domain and boundary conditions of ${\xi}$.
In this study, the thickness direction is represented by the $z$-axis, which is employed as a constraint to determine the presence or absence of the substrate. 
The anisotropic conductivities $\kappa_x$ and $\kappa_y$ are set to values substantially lower than $\kappa_z$. 
The boundary $\Gamma_{\xi}$, subjected to the Dirichlet boundary condition, is identified as the boundary opposite to domain $D_{pe}$ within the domain $D_{sb}$. 
The second equation of (\ref{eq:form_fict}) ensures that ${\xi} = \bar{{\xi}}_s$ within the substrate material domain $\Omega_{sb}$.
The physical field ${\xi}$ behaves as a fictitious heat within the fixed design domain $D$. 
In the domain $\Omega_{sb}$, it acts as a heat source with a temperature ${\xi}=\bar{{\xi}}_s$, and a cold heat bath with ${\xi}=-\bar{{\xi}}_0$ is applied at the boundary $\Gamma_{\xi}$.
Within the substrate material domain $\Omega_{sb}$, the fictitious heat ${\xi}$ is applied and
 propagates 
 in the $z$-direction towards the domain $D_{pe}$ when a material is present within $\Omega_{sb}$. 
Conversely, in the absence of a material  within $D_{sb}$, 
the boundary condition 
${\xi} = -\bar{\xi}_0$ 
is conveyed to the domain $D_{pe}$ owing to the lack of ${\xi}$ propagation in $\Omega_{sb}$.
Consequently, within the piezoelectric design domain $D_{pe}$, ${\xi}$ assumes positive values above the substrate and negative values elsewhere. 
By defining $\Omega_{pe}$ as regions where $\phi_{p\psi} > 0$ and ${\xi} > 0$, we ensure that the structure adheres to the substrate-dependent constraint:

\begin{align}
	&\chi_p (\bm{x}) = \left\{ 
	\begin{array}{ll}
    	1 \qquad \mathrm{for} \quad\bm{x}\in \Omega_{pe} \\
    	0 \qquad \mathrm{for} \quad\bm{x}\in D\backslash\Omega_{pe} , \\
	\end{array} 
	\right.\\
        &\chi_p(\bm{x}) := \chi_{\phi}(\phi_{p\psi}(\bm{x})) \chi_{\phi}(\xi(\bm{x}))
\end{align}

\begin{figure}[htpb]
    \centering
    \includegraphics[width=\linewidth]{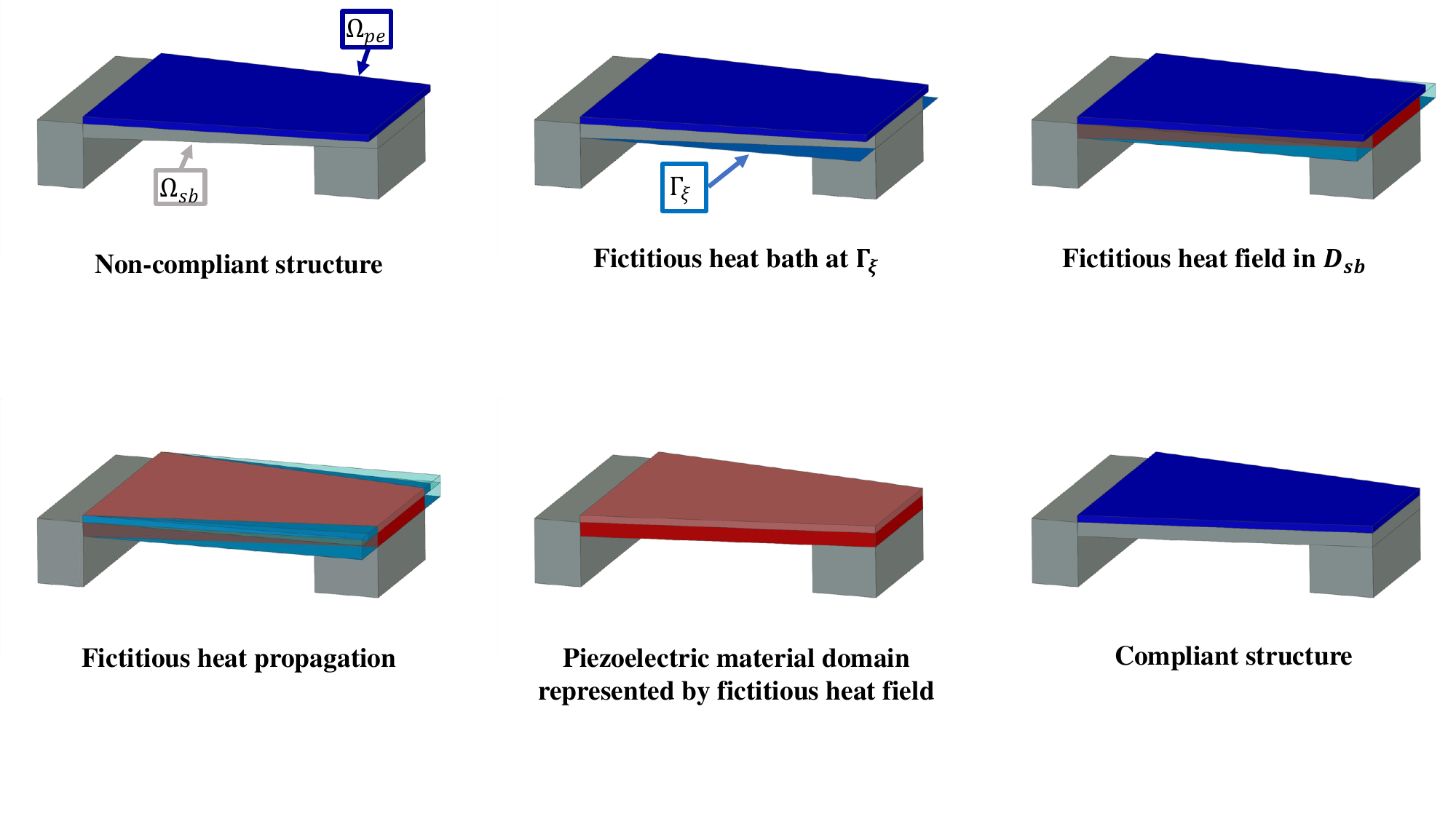} 
    \caption{
        Schematic illustration of the fictitious heat field used to enforce the substrate-dependent constraint.
        \textbf{Non-compliant Structure:} Initial configuration without constraints.
        \textbf{Fictitious Heat Bath at $\Gamma_{\xi}$:} Cold heat bath with $ \xi = -\bar{\xi}_0 $ is applied at $ \Gamma_{\xi} $, the boundary opposite to domain $ D_{pe} $ within domain $D_{sb}$.
        \textbf{Fictitious Heat Field in $D_{sb}$:} Substrate material domain $ \Omega_{sb} $ as a heat source with $ \xi = \bar{\xi}_s $.
        \textbf{Fictitious Heat Propagation:} Heat propagation in the $z$-direction towards $D_{pe}$. In the absence of material in $\Omega_{sb}$, $ \xi = -\bar{\xi}_0 $ is conveyed to $D_{pe}$.
        \textbf{Piezoelectric Domain Represented by Fictitious Heat Field:} $ \Omega_{pe} $ expressed as $\chi_p(\bm{x}) := \chi_{\phi}(\phi_{p\psi}(\bm{x})) \chi_{\phi}(\xi(\bm{x}))$ to satisfy the substrate-dependent constraint.
        \textbf{Compliant Structure:} Structure satisfying the substrate-dependent constraint.
    }
    \label{fig:fictitous}
\end{figure}

Figure \ref{fig:fictitous} visually elucidates the behavior of the fictitious heat field ${\xi}$ and its role in defining the piezoelectric material domain. 
The figure depicts the initial structure without constraints, followed by the propagation of the fictitious heat field within both the substrate and piezoelectric domains. 
The fictitious heat field ensures that no piezoelectric material is placed without the presence of a substrate.

%-----------------------------------------------------------
\subsection{Formulation of the optimal design with proposed constraints}
\label{subsec:form_opt}

By incorporating the aforementioned methodologies, the optimal design problem is formulated as follows:
\begin{align}
    \begin{split}
        \underset{{\phi_{p \psi}}}{\text{inf}}\qquad &F_{pe}\\
        \underset{{\phi_{s \psi}}}{\text{inf}}\qquad &F_{sb}\\
        \text{subject} \; \text{to} \qquad & \text{for} \; i = 1, \ldots, n \\
        &a_{oci}({\phi_{p \psi}}, {\phi_{s \psi}}, \omega_{oci},\bm{u}_{oci}, \bm{v}_{oci}, \varphi_{oci}, {\xi}) = 0,\\
        &b_{oci}({\phi_{p \psi}}, {\phi_{s \psi}}, \bm{u}_{oci}, \varphi_{oci}, v_{\varphi_{oci}}, {\xi}) = 0,\\
        &a_{sci}({\phi_{p \psi}}, {\phi_{s \psi}}, \omega_{sci},\bm{u}_{sci}, \bm{v}_{sci}, \xi) = 0,\\
        &a_{\xi}({\phi_{p \psi}}, {\phi_{s \psi}}, {\xi}, v_{\xi}) = 0,\\
        &G_V({\phi_{p \psi}}, \varphi_{n}) \le 0,
        \label{eq:form_piezo_sub_p}
    \end{split}
\end{align}

\begin{align}
    \begin{split}
        &a_{oci}({\phi_{p \psi}}, {\phi_{s \psi}}, \omega_{oci},\bm{u}_{oci}, \bm{v}_{oci}, \varphi_{oci}) \\
    	&\quad = \biggl( \int_{\Omega_{pe}} \bm{s}(\bm{v}_{oci})^T \bm{C}^E_{pe} \bm{s}(\bm{u}_{oci}) \dO 
    	   - \omega_{oc}^2  \int_{\Omega_{pe}} \rho_{pe} \bm{v}_{oci}^T \bm{u}_{oci} \dO 
    	   \\
    	  &\qquad +\int_{\Omega_{pe}} \bm{s} (\bm{v}_{oci})^T \bm{e}^T\bm{\nabla}\varphi_{oci} \dO 
    	  	%-\int_{\Omega_{pe}} \bm{v}^T \hat{\bm{f}} \dO
    	  \biggr)\\
    	  &\qquad+\biggl( 
    	  	 \int_{\Omega_{sb}} \bm{s}(\bm{v}_{oci})^T \bm{C}^E_{sb} \bm{s}(\bm{u}_{oci}) \dO 
    	  	- \omega_{oc}^2  \int_{\Omega_{sb}} \rho_{sb} \bm{v}_{oci}^T \bm{u}_{oci} \dO 
    	  	%-\int_{\Omega_{sb}} \bm{v}^T \hat{\bm{f}} \dO
    	  \biggr)=0,
    \end{split} \label{eq:def_aoc} \\
    \begin{split}
        &b_{oci}({\phi_{p \psi}}, {\phi_{s \psi}}, \bm{u}_{oci}, \varphi_{oci}, v_{\varphi_{oci}})\\
    	&\quad= \int_{\Omega_{pe}} \bm{\nabla}v_{\varphi_{oci}}^T \bm{e}\bm{s}(\bm{u}_{oci}) \dO 
    		- \int_{\Omega_{pe}} \bm{\nabla}v_{\varphi_{oci}}^T \bm{\varepsilon}^S\bm{\nabla}\varphi_{oci} \dO =0,
    \end{split} \label{eq:def_boc}\\
    \begin{split}
        &a_{sci}({\phi_{p \psi}}, {\phi_{s \psi}}, \omega_{sci},\bm{u}_{sci}, \bm{v}_{sci})\\
    	&\quad= \biggl(	
    		\int_{\Omega_{pe}} \bm{s}(\bm{v}_{sci})^T \bm{C}^E_{pe} \bm{s}(\bm{u}_{sci}) \dO 
    	    - \omega_{sci}^2  \int_{\Omega_{pe}} \rho_{pe} \bm{v}_{sci}^T \bm{u}_{sci} \dO
    	    %- \int_{\Omega_{pe}} \bm{v}^T \hat{\bm{f}} \dO
    	  \biggr)\\
    	  &\qquad+ \biggl(	
    	  		\int_{\Omega_{sb}} \bm{s}(\bm{v}_{sci})^T \bm{C}^E_{sb} \bm{s}(\bm{u}_{sci}) \dO 
    	  	    - \omega_{sci}^2  \int_{\Omega_{sb}} \rho_{sb} \bm{v}_{sci}^T \bm{u}_{sci} \dO
    	  	    %- \int_{\Omega_{sb}} \bm{v}^T \hat{\bm{f}} \dO
    	  	  \biggr) =0,
    \end{split} \label{eq:def_asc} \\
    \begin{split}
        &a_{\xi} ({\phi_{p \psi}}, {\phi_{s \psi}}, {\xi}, v_{\xi})= \int_{D} \nabla v_{\xi} \bm{\kappa}_{\xi}\nabla {\xi}  \dO + \int_{D} {h(\phi_{sp})} ({\xi}-\bar{{\xi}}_s(2\chi({\phi_{s \psi}}) - 1))v_{\xi} \dO =0,
    \end{split} \label{eq:def_ap} \\
    \begin{split}
        &G_V({\phi_{p \psi}}, \varphi_{n}) = \int_{\Omega_{pe}}\dO - \frac{\varepsilon_{z}  L_z^2 \int_{\Omega_{pe}}\bm{n}_z \cdot \nabla \varphi_{n} \dO}{\bar{V}_{minV}} \le 0,
    \end{split} \label{eq:def_Gv}
\end{align}
where $\rho_{pe}$ and $\rho_{sb}$ represent the densities of the piezoelectric and substrate materials, respectively, whereas $\bm{v}_{oci}$, $\bm{v}_{sci}$, $v_{\varphi oci}$, and $v_{\xi}$ serve as test functions. 
The test functions are detailed as follows:
$\bm{v}_{oci}$ and $\bm{v}_{sci}$ are test functions for the eigenvectors $\bm{u}_{oci}$ and $\bm{u}_{sci}$, respectively.
$v_{\varphi oci}$ is a test function for $\varphi_{oci}$.
$v_{\xi}$ is a test function for the fictitious physical function $\xi$.
The potential, symbolized by $\varphi_{n}$, is obtained from the displacement $\bm{u}$ derived via the modal superposition method. This is distinct from $\varphi_{oci}$, which is determined based on the eigenvector $\bm{u}_{oci}$.
A comprehensive exposition on the derivation of the potential $\varphi_{n}$ through the modal superposition method can be found in \ref{secA:solv}.

%-----------------------------------------------------------------------------------------------------------------------
\section{Numerical implementation}
\label{sec:impli}

%-----------------------------------------------------------
\subsection{Optimization algorithm}
The optimization algorithm is implemented as follows:
\begin{itemize}
	\item[\bf{Step 1:}] The initial level set function is set.
    \item[\bf{Step 2:}] The fictitious physical function field ${\xi}$ based on Eq. (\ref{eq:def_ap}) is solved using the finite element method (FEM).
	\item[\bf{Step 3:}] The eigenvector fields $\bm{u}_{oci}$ and $\bm{u}_{sci}$ and eigenvalues $\omega_{oci}$, $\omega_{sci}$, and $\varphi_{oci}$ derived from Eqs. (\ref{eq:def_aoc})--(\ref{eq:def_asc}) are solved using the FEM.
	\item[\bf{Step 4:}] {The displacement $\bm{u}$, the electric potential $\varphi_n$,} the objective functions $F_{pe}$ and $F_{sb}$ derived from Eq. (\ref{eq:form_piezo_ev_gen}) together with the output voltage $V_E$ at a frequency $\bar{\omega_i}$ and the constraint function $G_V$, which is based on Eq. (\ref{eq:def_Gv}), are evaluated.
	\item[\bf{Step 5:}] If the objective functions converge, the optimization procedure is terminated; otherwise, the sensitivity with respect to the objective functions is computed.
	\item[\bf{Step 6:}]The level set function is updated using the time evolution equation given by Eq. (\ref{eq:RDE}), and the optimization procedure returns to step 2.
\end{itemize}

%FreeFEM++ \cite{MR3043640} was used as the FEM solver.
{All computations, including those involving the FEM solver, are carried out using FreeFEM++ \cite{MR3043640}.
In this study, convergence is considered achieved when the ratio of the objective function value from the previous iteration to that of the current iteration remains below $1.0\times10^{-6}$ for 10 consecutive iterations. Additionally, the computation is terminated if the number of iterations reaches 1000.}
In the following subsections, we explain the approximations of the displacement  and electrical potential fields and the sensitivity analysis.

%-----------------------------------------------------------
\subsection{Approximate solution of the displacement and electrical potential fields based on the Eulerian coordinate system}
\label{subsec:euler}

Within the Eulerian coordinate system, the fixed design domain necessitates the generation of finite elements during each iteration of the optimization process.
To mitigate the computational expenditure, an ersatz material approach, as delineated in \cite{allaire2004structural}, is employed.
In particular, the void domain is assumed to be a structural material with a relatively small {Young's modulus}, and the material properties are assumed to be smoothly distributed in the vicinity of the interface {between the weak and strong phases}.
The governing equations, which dictate the displacement field within the FEM, originally pertain to the material domain $\Omega$.
Utilizing the level set, characteristic, and fictitious physical functions delineated in Section \ref{sec:constraint}, coupled with an approximated Heaviside function $h(\phi)$ to be presented subsequently, we extend these equations to the entire fixed design domain $D$ as follows:
\begin{align}
	\begin{split}
    	\chi_{\phi}(\phi) &\approx h(\phi) ,\\
    	h(\phi) :&= \left\{ 
		\begin{array}{lll}
    		d \quad &\mathrm{for} \quad \phi < -w  \\
    		\left\{\frac{1}{2}+\frac{\phi}{w} \left[ \frac{15}{16} - \frac{\phi^2}{w^2}(\frac{5}{8}-\frac{3}{16}\frac{\phi^2}{w^2}) \right]\right\}(1-d) + d \quad &\mathrm{for} \quad -w \leq \phi \leq w \\
    		1  \quad &\mathrm{for} \quad w<\phi \\
		\end{array} 
		,\right.
	\end{split}\\
    {\bm{C}_{\phi ps}^E} :&= \bm{C}^E_{pe}{h(\phi_{ps})}h({\phi_{p \psi}})h({\xi})+\bm{C}^E_{sb}{h(\phi_{sp})}h({\phi_{s \psi}}) , \label{eq:def_C_phi_s_p}\\
    {\bm{e}}_{\phi ps} :&= \bm{e}{h(\phi_{ps})}h({\phi_{p \psi}})h({\xi}) , \label{eq:def_e_phi_s_p}\\
    {\bm{\varepsilon}}^S_{\phi ps} :&= \varepsilon_0 \bm{I}(1-({h(\phi_{ps})}h({\phi_{p \psi}})h({\xi})+{h(\phi_{sp})}h({\phi_{s \psi}})))+ \bm{\varepsilon}^S{h(\phi_{ps})}h({\phi_{p \psi}})h({\xi}) , \label{eq:def_epsilon_phi_s_p}\\
    {\rho}_{\phi ps} :&= \rho_{pe} {h(\phi_{ps})}h({\phi_{p \psi}})h({\xi})+\rho_{sb}{h(\phi_{sp})}h({\phi_{s \psi}}),\label{eq:def_rho_phi_s_p}
\end{align}
where $\bm{C}^E_{\phi ps}$ is the extended elastic tensor, ${\bm{e}}_{\phi ps}$ is the extended piezoelectric constant tensor, ${\bm{\varepsilon}}^S_{\phi ps}$ is the extended permittivity tensor, ${\rho}_{\phi ps}$ is the extended density, $w$ is the transition width of the Heaviside function, and $d$ is a sufficiently small positive number.
In this study, we set $w=0.9$ and $d=0.01$.

%-----------------------------------------------------------
\subsection{Sensitivity analysis}
\label{subsec:sens}

Here, we describe a procedure for determining the sensitivity to update level set functions.
In this study, based on Eqs. (\ref{eq:def_aoc})--(\ref{eq:def_Gv}), we define the Lagrangian functions $L_{pe}$ and $L_{sb}$ of the level set functions with respect to ${\phi_{p \psi}}$ and ${\phi_{s \psi}}$, respectively, as follows:
\begin{align}
    \begin{split}
        \label{eq:Lpe}
        L_{pe} = &F_{pe} + 
        {\sum_{i=1}^n \biggl( }
        a_{oci}({\phi_{p \psi}}, {\phi_{s \psi}}, \omega_{oci},\bm{u}_{oci}, \bm{v}_{ocpe}, \varphi_{oci}) 
        + b_{oci}({\phi_{p \psi}}, {\phi_{s \psi}}, \bm{u}_{oci}, \varphi_{oci}, v_{\varphi_{ocpe}}) \\
        &+ a_{sci}({\phi_{p \psi}}, {\phi_{s \psi}}, \omega_{sci},\bm{u}_{sci}, \bm{v}_{scpe}) {\biggr)}
        + a_{\xi} ({\phi_{p \psi}}, {\phi_{s \psi}}, {\xi}, v_{\xi}) + \lambda G_V({\phi_{p \psi}}, \varphi_{n}),
    \end{split}\\
    \begin{split}
        \label{eq:Lsb}
        L_{sb} = &F_{sb} + {\sum_{i=1}^n \biggl(  }
        a_{oci}({\phi_{p \psi}}, {\phi_{s \psi}}, \omega_{oci},\bm{u}_{oci}, \bm{v}_{ocsb}, \varphi_{oci}) 
        + b_{oci}({\phi_{p \psi}}, {\phi_{s \psi}}, \bm{u}_{oci}, \varphi_{oci}, v_{\varphi_{ocsb}}) \\
        &+ a_{sci}({\phi_{p \psi}}, {\phi_{s \psi}}, \omega_{sci},\bm{u}_{sci}, \bm{v}_{scsb}) {\biggr)}
        + a_{\xi} ({\phi_{p \psi}}, {\phi_{s \psi}}, {\xi}, v_{\xi}),
    \end{split}
\end{align}
where $\bm{v}_{ocpe}$, $\bm{v}_{ocsb}$, $\bm{v}_{scpe}$, $\bm{v}_{scsb}$, $v_{\varphi_{ocpe}}$, $v_{\varphi_{ocsb}}$, and $\lambda$ are the Lagrange multipliers.
Using the adjoint method, sensitivity is determined as follows:
\begin{align}
    \begin{split}
        {F}'_{pe} = &{\sum_{i=1}^n \biggl(}c_{ocpe} \left\lbrace  a_{oci}(1, {\phi_{s \psi}}, \omega_{oci}, \bm{u}_{oci}, \bm{u}_{oci}, \varphi_{oci})
        +  b_{oci} (1, \bm{u}_{oci}, \varphi_{oci}, \varphi_{oci}) \right\rbrace \\
        &+ c_{scpe} a_{sci}(1, {\phi_{s \psi}}, \omega_{sci}, \bm{u}_{sci}, \bm{u}_{sci}) {\biggr)} + \lambda,
        \end{split}\\
        \begin{split}
        {F}'_{sb} =&{\sum_{i=1}^n \biggl(}c_{ocsb} \left\lbrace  a_{oci}({\phi_{p \psi}}, 1, \omega_{oci}, \bm{u}_{oci}, \bm{u}_{oci}, \varphi_{oci}) 
        \right\rbrace \\
        &+ c_{scsb} \left\lbrace a_{sci}({\phi_{p \psi}}, 1, \omega_{sci}, \bm{u}_{sci}, \bm{u}_{sci}) \right\rbrace {\biggr)}.
    \end{split}
\end{align}
The sensitivity analysis is explained in detail in \ref{secA:sens}.

%-----------------------------------------------------------------------------------------------------------------------
\section{Numerical examples}
\label{sec:example}
In this section, several numerical examples are presented to demonstrate the utility and validity of the proposed method.

\begin{figure}[htbp]
	\centering
	% Use the relevant command to insert your figure file.
	% For example, with the graphicx package use
	\includegraphics[scale=0.5]{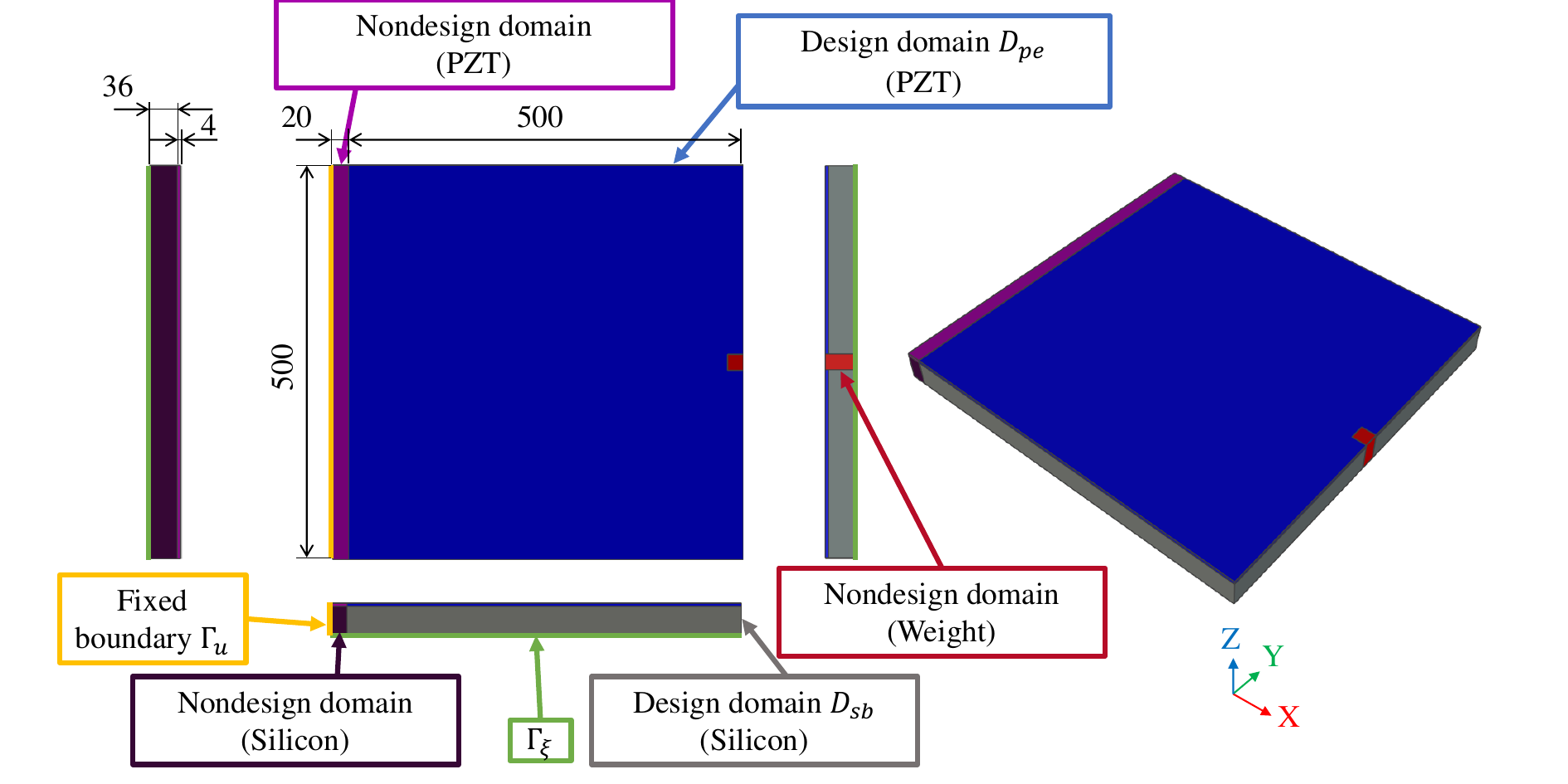}
	% figure caption is below the figure
	\caption{Design domain and boundary conditions.}
	\label{fig:domain1}       % Give a unique label
\end{figure}

The design domain and associated boundary conditions are illustrated in Figure \ref{fig:domain1}.
Let $D_{pe}$ denote a square, colored in blue in the figure, with a side length of 500 mm and a thickness of 4 mm.
The material in this domain is PZT, which is a piezoelectric material.
Similarly, let $D_{sb}$ symbolize a square, indicated in grey in the figure, with a side length of 500 mm and a thickness of 36 mm. 
The material within this domain is Silicon, which is a nonpiezoelectric material.
The fixed design domain $D$ consists of the combined regions of  $D_{pe}$ and $D_{sb}$.
The domain measuring 500 mm in length, 20 mm in width, and 36 mm in thickness, depicted in dark purple in the figure, is a 
nondesign 
domain composed of Silicon. 
This domain will be referred to as the 
"Nondesign domain (Silicon)".

Correspondingly, the domain that is 500 mm in length, 20 mm in breadth, and 4 mm in thickness, illustrated in light purple in the figure, represents a 
nondesign 
domain filled with PZT. 
This domain will be referred to as the 
"Nondesign domain (PZT)".

A fixed boundary condition is imposed on the surface of the 
Nondesign 
domain (Silicon) that is opposite to the surface adjacent to the domain $D_{sb}$.
The domain colored red on the right side of the figure, delineating a 20-mm square with a thickness of 40 mm, is designated as the weight domain, which is a 
nondesign 
domain filled with objects devoid of piezoelectric properties.
To reduce computational load, the weight part is represented as a 
nondesign 
domain with a high density.
The density is assumed to be 100 times that of the substrate material within this weight domain.
This domain will be referred to as the 
"Nondesign domain (Weight)".
The mesh configuration for different domains within this example is illustrated in Figure~\ref{fig:mesh_config}. 
In the 
nondesign 
domain (PZT), the mesh size is uniformly set to 10 $\times$ 10 $\times$ 2. 
The design domain $D_{pe}$ (PZT) also has a uniform mesh size of 20 $\times$ 10 $\times$ 2. 
For the 
nondesign 
domain (Silicon), the mesh size is 20 $\times$ 10 $\times$ 8.5 for $z$ coordinates less than 34, and 20 $\times$ 10 $\times$ 2 for $z$ coordinates 34 and above, near the interface with $D_{pe}$. 
The 
nondesign 
domain (Weight) follows the same pattern, with a mesh size of 20 $\times$ 10 $\times$ 8.5 for $z$ coordinates less than 34, and 20 $\times$ 10 $\times$ 2 for $z$ coordinates 34 and above. 
Finally, the design domain $D_{sb}$ (Silicon) has a mesh size of 20 $\times$ 10 $\times$ 8.5 for $z$ coordinates less than 34, except at the interface with $D_{pe}$, where the mesh size is 20 $\times$ 10 $\times$ 2 for $z$ coordinates 34 and above.
For each condition, the computation time took approximately 100 hours.
The number of modes to be contemplated is fixed at $n=4$, and the specified natural frequencies are $[\bar{\omega}_{0}, \bar{\omega}_{1}, \bar{\omega}_{2}, \bar{\omega}_{3}] = [70 \;\text{Hz}, 435 \;\text{Hz}, 450 \;\text{Hz}, 500 \;\text{Hz}]$.
The initial structure is conjectured to be such that ${\phi_{p \psi}} = 1$ in $D_{pe}$ and ${\phi_{s \psi}} = 1$ in $D_{sb}$, indicating that the entire fixed design domain is occupied by the material domain.

{
\begin{figure}[htbp]
    \centering
    \includegraphics[scale=0.7]{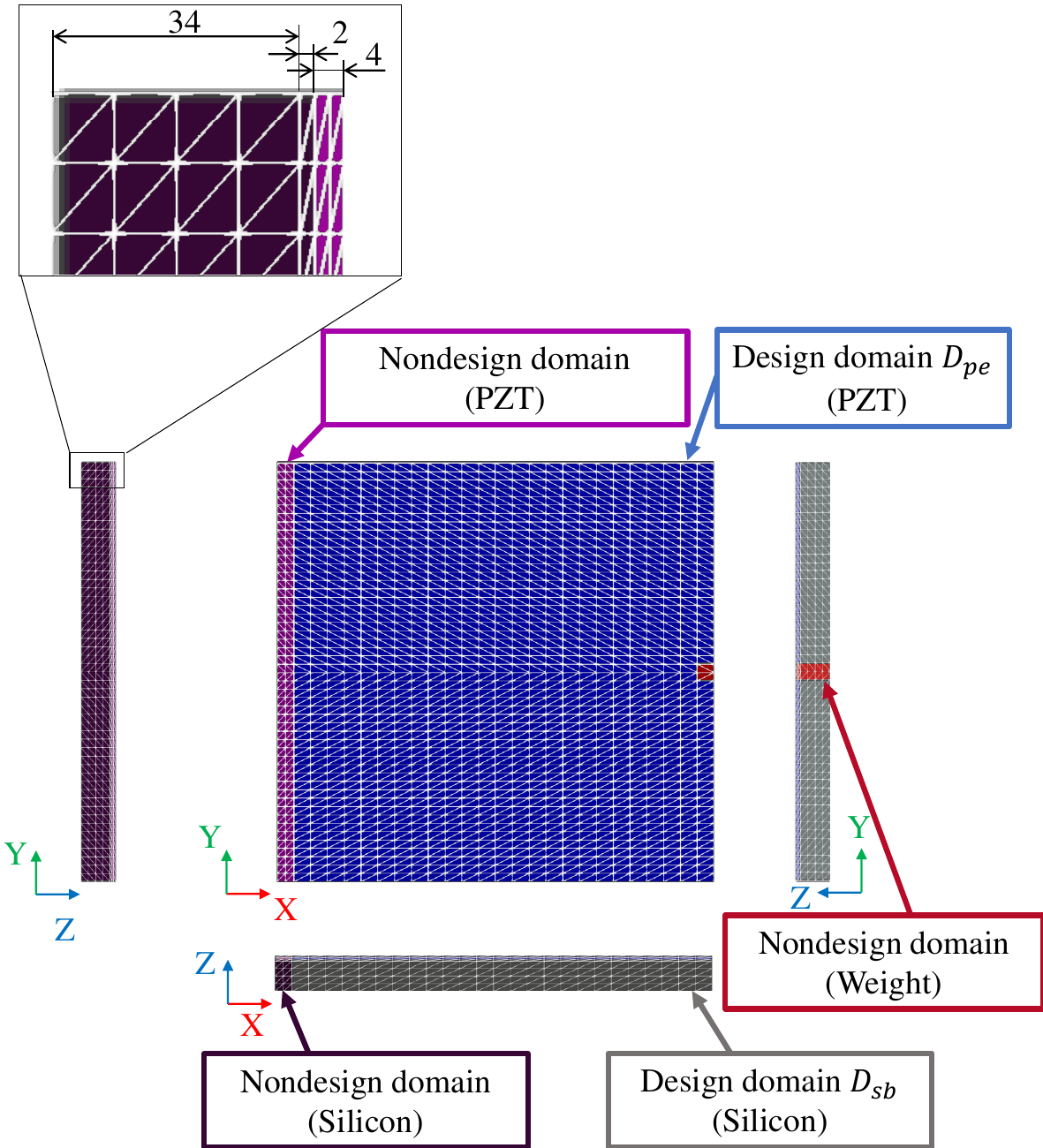}
    \caption{Mesh configuration in different domains. In the 
    nondesign 
    domain (PZT), the mesh size is uniformly set to 10 $\times$ 10 $\times$ 2. The design domain $D_{pe}$ (PZT) has a uniform mesh size of 20 $\times$ 10 $\times$ 2. For the 
    nondesign 
    domain (Silicon), the mesh size is 20 $\times$ 10 $\times$ 8.5 for $z$ coordinates less than 34, and 20 $\times$ 10 $\times$ 2 for $z$ coordinates 34 and above, near the interface with $D_{pe}$. The 
    nondesign 
    domain (Weight) has a mesh size of 20 $\times$ 10 $\times$ 8.5 for $z$ coordinates less than 34, and 20 $\times$ 10 $\times$ 2 for $z$ coordinates 34 and above. The design domain $D_{sb}$ (Silicon) has a mesh size of 20 $\times$ 10 $\times$ 8.5 for $z$ coordinates less than 34, except at the interface with $D_{pe}$, where the mesh size is 20 $\times$ 10 $\times$ 2 for $z$ coordinates 34 and above.}
    \label{fig:mesh_config}
\end{figure}
}
%-----------------------------------------------------------
\subsection{Effectiveness of the proposed constraints for manufacturability}
\label{subsec:num_ex_1}

In this subsection, we present numerical examples to confirm the effectiveness of the constraints for manufacturability proposed in Section \ref{sec:constraint}.
In order to verify the effect of "Constraint: the same cross-sectional shape throughout the design domain" mentioned in subsection \ref{subsec:tau}, we test the regularization factor $\tau$.
Similarly, to verify the effect of "Constraint: the same cross-sectional shape in each domain" discussed in subsection \ref{subsec:ex-ls}, we examine the representation of level set functions.
Lastly, to verify the effect of "Constraint: no piezoelectric material is placed without a substrate" described in subsection \ref{subsec:p}, we evaluate the status of constraint implementation via a fictitious field ${\xi}$.

\begin{figure}[htbp]
	\centering
	% Use the relevant command to insert your figure file.
	% For example, with the graphicx package use
	\includegraphics[scale=0.6]{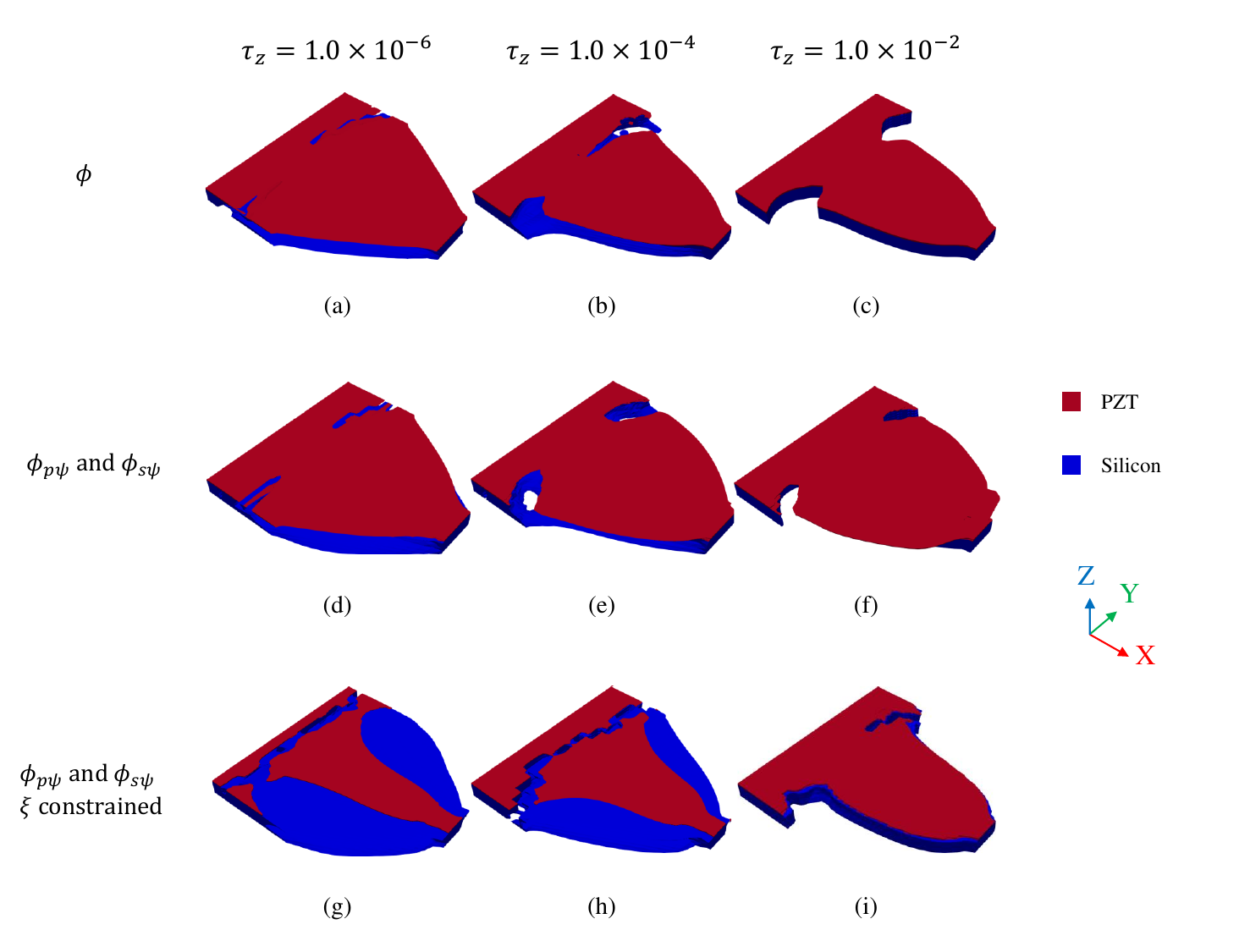}
	% figure caption is below the figure
	\caption{Optimal configuration of a 
                unimorph cantilevered energy harvester for conditions (a)--(i) listed in Table \ref{tab:num_ex_cond}. The red region denotes the piezoelectric material domain $\Omega_{pe}$, and the blue region represents the substrate material domain $\Omega_{sb}$.}
	\label{fig:shape_num_ex}       % Give a unique label
\end{figure}

We conduct numerical experiments under nine conditions, as listed in Table \ref{tab:num_ex_cond}.
Each condition is characterized by a distinct setting of the regularization factor $\tau_z$ in the thickness direction, the representation of level set functions either a singular function $\phi$ or a pair of functions ${\phi_{p \psi}}$ and ${\phi_{s \psi}}$, and the status of constraint implementation via a fictitious field ${\xi}$. 
Specifically, for each condition, the regularization factor $\tau_z$ is set to 1.0$\times 10^{-6}$, 1.0$\times 10^{-4}$, or 1.0$\times 10^{-2}$. 
Furthermore, the level set functions are either represented solely by $\phi$ or by a combination of ${\phi_{p \psi}}$ and ${\phi_{s \psi}}$, and the constraints are applied using the fictitious field ${\xi}$ for conditions (g)--(i).
Each of these nine configurations is evaluated through 1000 iterations of computation.
The parameters corresponding to each condition, coupled with the average values of the objective function $F_k$ and $F_{\omega}$ taken over the last 100 iterations, are also presented in Table \ref{tab:num_ex_cond}.
Under all conditions, the weighting factors $\alpha_{pe}$ and $\alpha_{sb}$ are set to $0.95$ to achieve a balanced contribution between $F_k$ and $F_{\omega}$ in both the piezoelectric and the substrate domains.
The resulting structures for each condition are shown in Figure \ref{fig:shape_num_ex}.

\begin{figure}[htbp]
	\centering
	% Use the relevant command to insert your figure file.
	% For example, with the graphicx package use
	\includegraphics[scale=0.6]{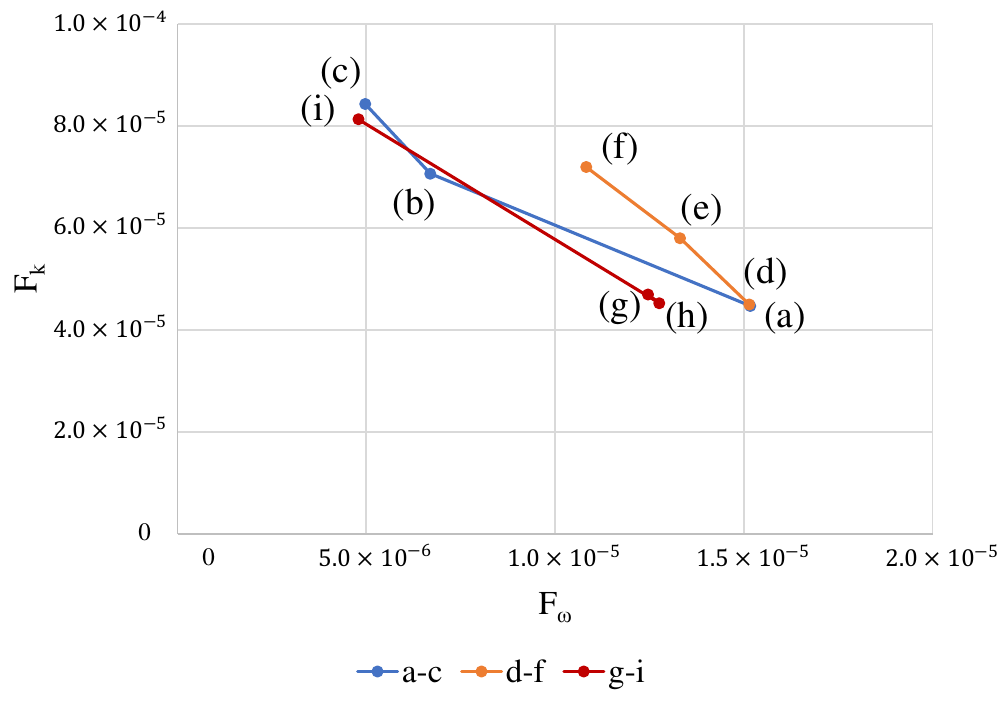}
	% figure caption is below the figure
	\caption{Tradeoff relationship between the objective functions $F_k$ and $F_{\omega}$ under conditions (a)--(i) listed in Table \ref{tab:num_ex_cond}.}
	\label{fig:fk_Fo}       % Give a unique label
\end{figure}

The tradeoff relationship between the objective functions $F_k$ and $F_{\omega}$ for conditions (a)--(i) delineated in Table \ref{tab:num_ex_cond} is depicted in Figure \ref{fig:fk_Fo}. 
From the figure, a clear tradeoff relationship can be discerned between $F_\omega$ and $F_k$, the balance of which is influenced by changes in the parameter $\tau_z$. 
Moreover, conditions (d)--(f) yield smaller $F_k$ values than conditions (a)--(c), given the same $\tau_z$ value. 
This phenomenon is attributed to the fact that, in conditions (d)--(f), the architecture of the piezoelectric film remains uninfluenced by the structure of the substrate, and, apart from regularization with respect to ${\phi_{p \psi}}$, it does not adhere to any constraints.
This allows for a wider variety of potential solutions, thereby enhancing the structural degrees of freedom of piezoelectric films. 
In conditions (g)--(i), although the structure of the piezoelectric film is influenced by the structure of the substrate, the converse does not hold true, resulting in $F_\omega$ values that are lower than those under conditions (d)--(f).

\begin{table}[tbp]
 \caption{Summary of parameter settings and objective functions: regularization factor $\tau_{z}$ in the thickness direction, the representation of level set functions, and the status of constraint implementation via a fictitious field ${\xi}$ and the corresponding objective functions $F_k$ and $F_{\omega}$ for each condition.}
 \label{tab:num_ex_cond}
 \centering
 \begin{tabular}{cccccc}
     \hline
     & & & &  \multicolumn{2}{c}{Objective function}\\
      condition & $\tau_z$ & level set function & ${\xi}$ constraint &$F_k$&$F_{\omega}$\\
     \hline
     (a) &1.0$\times 10^{-6} $& $\phi$& not constrained&4.47$\times 10^{-5}$&1.52$\times 10^{-5}$\\
     (b) &1.0$\times 10^{-4} $& $\phi$& not constrained&7.07$\times 10^{-5}$&6.70$\times 10^{-6}$\\
     (c) &1.0$\times 10^{-2} $& $\phi$& not constrained&8.43$\times 10^{-5}$&4.98$\times 10^{-6}$\\
     (d) &1.0$\times 10^{-6} $& ${\phi_{p \psi}} \;$ and $\; {\phi_{s \psi}}$& not constrained&4.50$\times 10^{-5}$&1.51$\times 10^{-5}$\\
     (e) &1.0$\times 10^{-4} $& ${\phi_{p \psi}} \;$ and $\; {\phi_{s \psi}}$& not constrained&5.80$\times 10^{-5}$&1.33$\times 10^{-5}$\\
     (f) &1.0$\times 10^{-2} $& ${\phi_{p \psi}} \;$ and $\; {\phi_{s \psi}}$& not constrained&7.20$\times 10^{-5}$&1.08$\times 10^{-5}$\\
     (g) &1.0$\times 10^{-6} $& ${\phi_{p \psi}} \;$ and $\; {\phi_{s \psi}}$& ${\xi}$ constrained&4.70$\times 10^{-5}$&1.25$\times 10^{-5}$\\
     (h) &1.0$\times 10^{-4} $& ${\phi_{p \psi}} \;$ and $\; {\phi_{s \psi}}$& ${\xi}$ constrained&4.53$\times 10^{-5}$&1.28$\times 10^{-5}$\\
     (i) &1.0$\times 10^{-2} $& ${\phi_{p \psi}} \;$ and $\; {\phi_{s \psi}}$& ${\xi}$ constrained&8.14$\times 10^{-5}$&4.79$\times 10^{-6}$\\
     \hline
 \end{tabular}
\end{table}

Next, the following paragraph explains the effectiveness of the proposed constraints.
We compute two normalized quantities: $N_{\phi1}$ and $N_{\phi2}$.
$N_{\phi1}$ is calculated to confirm the extent to which boundaries in the $z$ direction exist within the entire design domain $D$.
This $N_{\phi1}$ verifies whether the constant cross-sectional constraint is effective throughout the fixed design domain $D$.
$N_{\phi1}$ is the proportion of nodes where the level set function changes sign with adjacent nodes in the negative $z$-direction, indicating a boundary.
Conversely, $N_{\phi2}$ is the proportion of nodes
, among those extracted by $N_{\phi1}$, that are 
not located at the boundaries of domains $D_{sb}$ and $D_{pe}$.
By calculating $N_{\phi2}$, we verify whether the constraint of having a constant cross-section in each of $D_{pe}$ and $D_{sb}$ is effective.
These values are normalized by the total number of nodes.
The computed values for $N_{\phi1}$ and $N_{\phi2}$ for each condition are provided in Table \ref{tab:N_phi}.

As delineated in Table \ref{tab:N_phi}, irrespective of the level set function configuration and the existence of a ${\xi}$ constraint, an increase in the $\tau_z$ value corresponds to a reduction in the z-directional complexity.
When two level set functions are utilized, as in conditions (d)--(i), the $N_{\phi1}$ value surpasses that in conditions (a)--(c), where a single level set function embodied the entirety of the design domain.
Contrastingly, $N_{\phi2}$, excluding the interface between $D_{sb}$ and $D_{pe}$, does not exhibit a significant difference from $N_{\phi1}$ under conditions (a)--(c) but is less than $N_{\phi1}$ in conditions (d)--(i).
This phenomenon can be attributed to the fact that, when two level set functions are employed, the structural complexity is not restrained at the boundary between domains $D_{sb}$ and $D_{pe}$, even upon altering the $\tau_z$ value. 
Hence, the $N_{\phi1}$ value is considered higher than that in conditions (a)--(c).

\begin{table}[htbp]
 \caption{Number of nodes where the product of their level set function values with nodes adjacent to the negative direction of the z-axis is negative for each condition}
 \label{tab:N_phi}
 \centering
 \begin{tabular}{cccc}
     \hline
      condition & $\tau_z$ & $N_{\phi1}$&$N_{\phi2}$\\
     \hline
     (a) &1.0$\times 10^{-6} $& 0.08818 & 0.07629\\
     (b) &1.0$\times 10^{-4} $& 0.01965 & 0.01819\\
     (c) &1.0$\times 10^{-2} $& 0.00049 & 0.00042\\
     (d) &1.0$\times 10^{-6} $& 0.08890 & 0.07138\\
     (e) &1.0$\times 10^{-4} $& 0.05058 & 0.03523\\
     (f) &1.0$\times 10^{-2} $& 0.01898 & 0.00079\\
     (g) &1.0$\times 10^{-6} $& 0.14124 & 0.10782\\
     (h) &1.0$\times 10^{-4} $& 0.08660 & 0.06932\\
     (i) &1.0$\times 10^{-2} $& 0.00340 & 0.00200\\
     \hline
 \end{tabular}
\end{table}

%-----------------------------------------------------------
\subsection{Optimal configuration subjected to a minimum output voltage constraint}

\begin{figure}[htbp]
	\centering
	% Use the relevant command to insert your figure file.
	% For example, with the graphicx package use
	\includegraphics[scale=0.4]{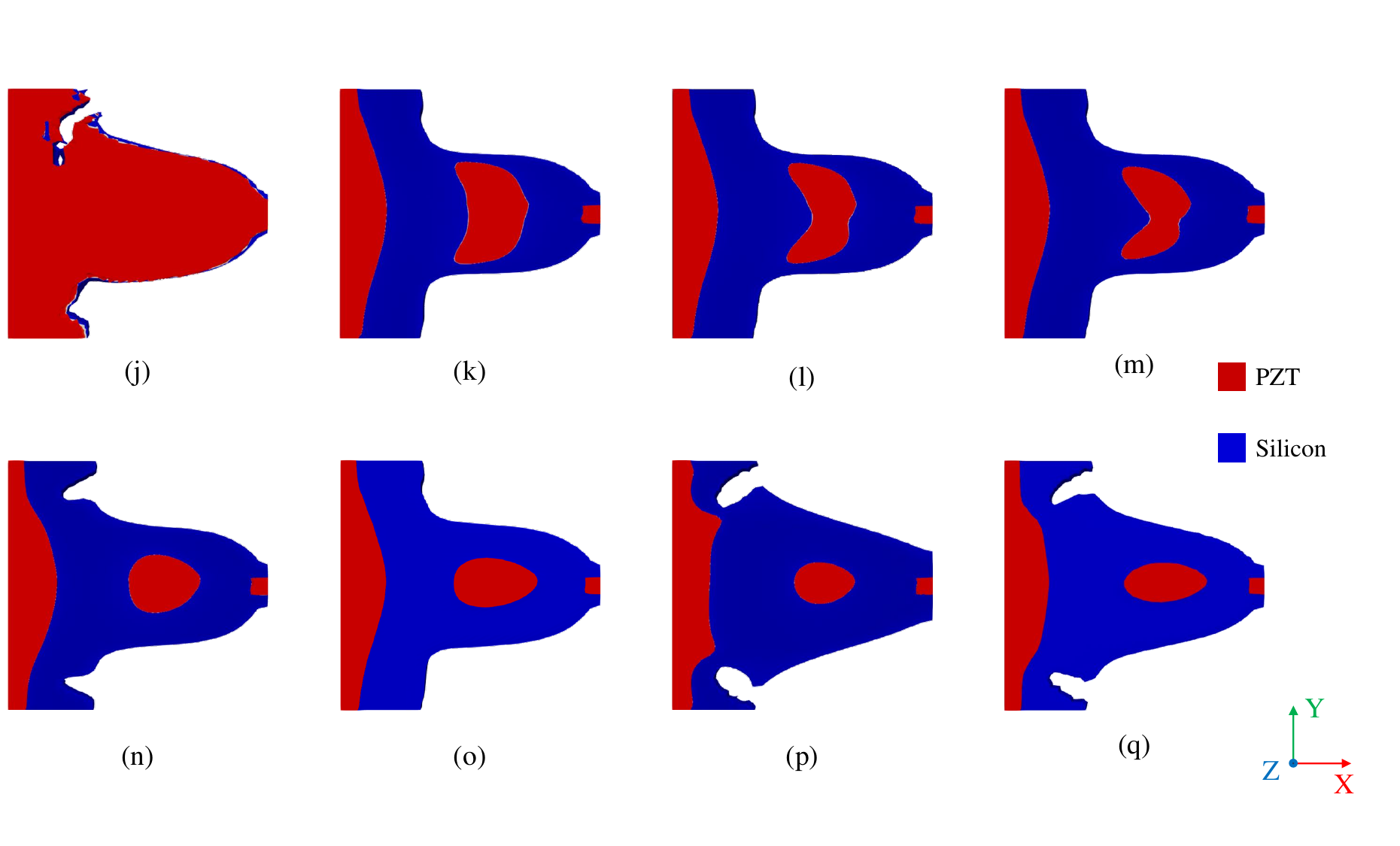}
	% figure caption is below the figure
	\caption{Optimal configuration of the unimorph cantilevered energy harvester for conditions (j)--(q), as detailed in Table \ref{tab:V_const_cond}. The red region denotes the piezoelectric material domain $\Omega_{pe}$, and the blue region represents the substrate material domain $\Omega_{sb}$.}
	\label{fig:shape_vconst}       % Give a unique label
\end{figure}

Herein, the efficacy of the introduced constraint with respect to output voltage is critically examined.
Consistently across all scenarios, the substrate-dependent constraint is enforced utilizing the fictitious physical function ${\xi}$.
Moreover, the material domain is characterized by two level set functions, ${\phi_{p \psi}}$ and ${\phi_{s \psi}}$, with the parameter $\tau_z$ assigned a value of $1.0 \times 10^{-2}$. 
Table \ref{tab:V_const_cond} enumerates the parameters for the minimum output voltage constraint under various conditions, alongside their respective objective functions $F_k$, $F_\omega$, and output voltage $V_E$. 
The optimal configurations for each scenario are graphically represented in Figure \ref{fig:shape_vconst}.

\begin{table}[btp]
 \caption{Summary of constraint parameters and objective functions: minimum output voltage constraint $\bar{V}_{minV}$ and the corresponding objective functions $F_k$ and $F_\omega$, the output voltage $V_E$, and the volume of piezoelectric material domain $\Omega_{pe}$ for each condition.}
 \label{tab:V_const_cond}
 \centering
 \begin{tabular}{cccccc}
     \hline
     & Constraint & \multicolumn{2}{c}{Objective function}& Output voltage& Volume of $\Omega_{pe}$\\ 
      condition & $\bar{V}_{minV}$&$F_k$&$F_\omega$&$V_E$ & $V_{pe}$\\
     \hline
     (j) &9.0$\times 10^{-3} $&8.44$\times 10^{-5}$  &4.56$\times 10^{-6}$& 9.73$\times 10^{-3}$ & 4.47 $\times 10^{5}$ \\
     (k) &9.5$\times 10^{-3} $&1.82$\times 10^{-4}$  &4.87$\times 10^{-6}$& 9.50$\times 10^{-3}$ & 2.31 $\times 10^{5}$  \\
     (l) &1.0$\times 10^{-2} $&2.03$\times 10^{-4}$  &5.60$\times 10^{-6}$& 1.00$\times 10^{-2}$ & 2.07 $\times 10^{5}$ \\
     (m) &1.05$\times 10^{-2} $&2.21$\times 10^{-4}$ &6.23$\times 10^{-6}$& 1.05$\times 10^{-2}$ & 1.90 $\times 10^{5}$\\
     (n) &1.1$\times 10^{-2} $&2.34$\times 10^{-4}$ &6.72$\times 10^{-6}$& 1.10$\times 10^{-2}$ & 1.80 $\times 10^{5}$\\
     (o) &1.15$\times 10^{-2} $&2.45$\times 10^{-4}$ &7.11$\times 10^{-6}$& 1.15$\times 10^{-2}$ & 1.71 $\times 10^{5}$\\
     (p) &1.2$\times 10^{-2} $&2.39$\times 10^{-4}$ &1.02$\times 10^{-5}$& 1.20$\times 10^{-2}$ & 1.66 $\times 10^{5}$\\
     (q) &1.3$\times 10^{-2} $&2.61$\times 10^{-4}$ &1.11$\times 10^{-5}$& 1.30$\times 10^{-2}$ & 1.49 $\times 10^{5}$\\
     \hline
 \end{tabular}
\end{table}

As shown in Table \ref{tab:V_const_cond}, the output voltage $V_E$ obtained under condition (j) 
exceeds 
the stipulated minimum voltage constraint. 
However, under the other conditions, the output voltages are aligned exactly with their respective constraints, highlighting the effectiveness and precision of the minimum output voltage constraint.
Furthermore, as the output voltage constraint values increase, there is a concomitant reduction in the volume of the piezoelectric material domain, represented by $\Omega_{pe}$. 
As elucidated by Eq. (\ref{eq:output_voltage_const}), this reduction in the volume of the piezoelectric material domain results in a diminished capacitance, which, in turn, amplifies the production voltage.
Concomitantly, an increase in the $\bar{V}_{minV}$ value invariably leads to a corresponding increase in both objective functions $F_k$ and $F_\omega$. 
It is cogent to infer that imposing a minimum output voltage constraint invariably curtails the potential for optimizing the objective functions $F_k$ and $F_\omega$.

As shown in Figure \ref{fig:shape_vconst}, under conditions (k)--(q) with an active minimum output voltage constraint, the piezoelectric material domain is predominantly located near the fixed boundary and at the midpoint of the beam.
Close to the fixed boundary of the beam, the aggregation of the piezoelectric material domain can be attributed to the stress concentration observed in the first-order eigenmode.
The region encompassing the piezoelectric material domain located around the  center of the beam is aligned with the zone of maximal displacement in the second-order eigenmodes.

%-----------------------------------------------------------------------------------------------------------------------
\section{Conclusion}
\label{sec:conclusion}

In this study, we propose a method for designing piezoelectric energy harvesters while improving their manufacturability.
The proposed method derives both the substrate and piezoelectric designs that maximize the electromechanical coupling coefficient. 
This method allows the eigenfrequency of the device and minimum output voltage to be set to the desired values.
Furthermore, the proposed method yields a design that can be manufactured using a  microfabrication process.
The results of this study are summarized as follows:

\begin{itemize}
	\item[1.] A topology optimization method is formulated to achieve the desired values for the eigenfrequency and minimum output voltage while concurrently maximizing the electromechanical coupling coefficient. An optimization algorithm is constructed for the numerical analysis.
	\item[2.] Cross-sectional shape constraints and substrate-dependent constraints are formulated to improve manufacturability with microfabrication. 
    \item[3.] Several numerical examples are provided to demonstrate the utility and validity of the proposed method. 
    The proposed method can provide solutions that satisfy the cross-sectional shape, substrate-dependent, and minimum output voltage constraints. 
\end{itemize}

\subsection{Limitations and future work} 
Despite the promising results, this study has some limitations. As the constraints become more stringent, the value of the objective function deteriorates. This indicates that the proposed method may struggle to provide solutions with significant electrical-mechanical coupling coefficients under highly restrictive conditions.

In future work, we aim to actually manufacture devices based on the designs obtained through the proposed method. This will allow us to validate the practicality of the solutions obtained and further demonstrate the utility of the proposed method. 

\subsection{Final remarks} 
The proposed method for designing piezoelectric energy harvesters has shown significant potential for improving their manufacturability. We believe that this research contributes to the field of energy harvesting and opens up new possibilities for the design of efficient and manufacturable energy harvesters.

%\section*{Acknowledgments}
%This work was partly supported by JSPS KAKENHI Grant Number 19H02049 and Katsu start-up fund of the University of Tokyo.
%------------------------------------------------------------------------------------------------------------------------
\appendix
\section{Computation of the output voltage using the modal superposition method}
\label{secA:solv}

In this study, the output voltage is computed using the modal superposition method.
In \ref{subsecA:eigen}, we explain the method used to solve the eigenvalue problem using the finite element method (FEM).
In \ref{subsecA:displacement}, we explain the method used to yield the output voltage using the modal superposition method based on the results obtained in \ref{subsecA:eigen}.

%-----------------------------------------------------------
\subsection{Eigenvalue and eigenvector determination}
\label{subsecA:eigen}

In this subsection, we elucidate an approach to solve the eigenvalue problem for piezoelectric materials using the FEM. 
Under open-circuit conditions, the relationships can be expressed in the form of finite element equations, as described below {\cite{erturk2011piezoelectric,SALAS2018223}}.
\begin{align}
    (\bm{K}  - \omega_{oc}^2\bm{M}) \bm{u} + \bm{P} \varphi &= 0,\\
    \bm{P}^T \bm{u} - \bm{G} \varphi&= 0,
\end{align}
wherein $\bm{u}$ and $\varphi$ correspond to the displacement and electric potential under open-circuit conditions, respectively
, 
and $\bm{P}$ and $\bm{G}$ denote the piezoelectric and dielectric matrices, respectively.
In the absence of an external charge, denoted by $\bm{q}$, and an external force, symbolized by $\bm{f}$, the potential $\varphi$ can be eliminated. 
This leads to:
\begin{align}
    \begin{split}
        &\bm{P}^T \bm{u} = \bm{G} \varphi ,\\
        &\varphi = \bm{G}^{-1} \bm{P}^T \bm{u} ,
        \label{eq:phi_oc}
    \end{split}\\
&(\bm{K}  - \omega^2\bm{M} + \bm{PG}^{-1}\bm{P}^T ) \bm{u} = 0 .
\end{align}
Consequently, the eigenvalue equation pertinent to the open-circuit conditions can be expressed as follows:
\begin{equation}
    \bm{M}^{-1}(\bm{K}  + \bm{PG}^{-1}\bm{P}^T ) \bm{u}_{oc} = \omega_{oc}^2 \bm{u}_{oc} ,
\end{equation}
where $\bm{u}_{oc}$ represents the eigenvector associated with the eigenvalue  $\omega_{oc}$.

Conversely, for a short-circuited electrode, the potential difference between the electrodes is zero.
By postulating that both the external charge $\bm{q}$ and the external force $\bm{f}$ are zero, analogous to the open-circuit condition, and by grounding the electrode to establish a zero potential, Eq. (\ref{eq:strong_short}) yields the following equation:
\begin{align}
    (\bm{K} - \omega^2\bm{M} ) \bm{u} = 0 .
\end{align}
The eigenvalue equation for the short-circuit condition is given as follows:
\begin{align}
    \bm{M}^{-1}\bm{K} \bm{u}_{sc} = \omega_{sc}^2 \bm{u}_{sc} ,
\end{align}
where $\bm{u}_{sc}$ represents the eigenvector corresponding to the eigenvalue $\omega_{sc}$.

%-----------------------------------------------------------
\subsection{Evaluation of displacement and output voltage using the modal superposition method}
\label{subsecA:displacement}

{In this subsection, we describe the process of obtaining the displacement field using the modal superposition method, utilizing the eigenvectors obtained in the previous subsection {\cite{erturk2011piezoelectric}}.}
To evaluate the output voltage, it is necessary to determine the displacement field. 
Since the eigenvectors are normalized, we need to consider the magnitude of the response for each mode to accurately determine the overall displacement.
Therefore, we employ the modal superposition method to determine the displacement field necessary for this evaluation.

To obtain the displacement field under forced vibration conditions, we initially express the equation of motion as:
\begin{align}
    \bm{M} \ddot{\bm{u}} + \bm{C} \dot{\bm{u}} + \bm{K}\bm{u} = \bm{f} .
    \label{eq:motion}
\end{align}

In the previous subsection, we discussed the general eigenvalue problem and denoted the solutions as $\bm{u}_{oc}$, $\omega_{oc}$, $\bm{u}_{sc}$, and $\omega_{sc}$.
Using the modal transformation matrix $\bm{{\Lambda}}$, we perform a coordinate transformation such that $\bm{u}=\bm{{\Lambda}} \bm{q}$. 
The modal transformation matrix $\bm{{\Lambda}}$ {and the modal amplitude vector $\bm{q}$ are} given by:
\begin{align}
    \bm{{\Lambda}} =& \left[ \bm{u}_{oc1}, \bm{u}_{oc2}, \bm{u}_{oc3} , \cdots , \bm{u}_{ocn}  \right] ,\\
    {
    \bm{q} =}&{ \left[ {q}_{1}, {q}_{2}, {q}_{3} , \cdots , {q}_{n}  \right].
    }
\end{align}

Multiplying Eq. (\ref{eq:motion}) on the left with $\bm{{\Lambda}}^T$, we obtain:
\begin{align}
    \bm{{\Lambda}}^T\bm{M}\bm{{\Lambda}}\ddot{\bm{q}}
    +\bm{{\Lambda}}^T\bm{C}\bm{{\Lambda}}\dot{\bm{q}}
    +\bm{{\Lambda}}^T\bm{K}\bm{{\Lambda}}\bm{q}
    =\bm{{\Lambda}}^T\bm{f} .
\end{align}
Through normalization, we transform each coefficient term as follows:
\begin{align}
    \bm{{\Lambda}}^T\bm{M}\bm{{\Lambda}} &= \bm{I},\\
    \bm{{\Lambda}}^T\bm{C}\bm{{\Lambda}} = \bm{D} &=
    \left[
    \begin{array}{llll}
        2\zeta_1 \omega_1 & 0 & \cdots & 0\\
        0 & 2\zeta_2 \omega_2& \cdots & 0\\
        \vdots & \vdots & \ddots & \vdots\\
        0 & 0 & \cdots & 2\zeta_n \omega_n
    \end{array}
    \right] ,\\
    \bm{{\Lambda}}^T\bm{K}\bm{{\Lambda}} &= \left[  
    \begin{array}{llll}
        \omega_1^2 & 0 & \cdots & 0\\
        0 & \omega_2^2 & \cdots & 0\\
        \vdots & \vdots & \ddots & \vdots\\
        0 & 0 & \cdots &  \omega_n^2
    \end{array}
    \right] ,\\
    \bm{{\Lambda}}^T\bm{f} &= \left[  
    \begin{array}{llll}
        F_1(t) & 0 & \cdots & 0\\
        0 & F_2(t) & \cdots & 0\\
        \vdots & \vdots & \ddots & \vdots\\
        0 & 0 & \cdots &  F_n(t)
    \end{array}
    \right] .
\end{align}
where $\zeta_i$ is the modal damping ratio for the $i$th mode.
In this transformation, we assume that the damping is small and the off-diagonal terms of the damping matrix are negligible compared with the diagonal terms. 
This transformation allows us to express the motion Eq. (\ref{eq:motion}) as $n$ independent equations
, one 
for each eigenmode as follows:
\begin{align}
    \ddot{q}_i(t) + 2\zeta_i \omega_i \dot{q}_i + \omega_i^2 q_i(t) = F_i(t)\qquad (i = 1,2, \cdots ,n),
\end{align}

\begin{align}
    q_i (t) = \frac{F_i(t)}{\omega_i\sqrt{\left( 1-(\bar{\omega}_i /\omega_i)^2\right)^2 +4\zeta_i^2(\bar{\omega}_i /\omega_i)^2 } }.
\end{align}
From the above, 
%displacement is yielded as follows:
{ the displacement is obtained as:}
\begin{align}
    \bm{u}(t) = \sum_{i=1}^n  q_i(t) \bm{u}_{oci}.
\end{align}

Incorporating this $\bm{u}$ into Eq. (\ref{eq:phi_oc}) yields the potential $\varphi_{n}$.
Notably, $\zeta_i = 0.01$ for $i= 1, 2, \cdots n$.
Finally, incorporating $\varphi_{n}$ into Eq. (\ref{eq:def_voltage}) yields the output voltage.

%--------------------------------------------------------------
\section{Sensitivity analysis}
\label{secA:sens}

In this section, we provide a detailed description of the sensitivity analysis using mathematical formulations.

At the stationary point of the Lagrangian, the following optimality conditions hold:
\begin{empheq}[left=\empheqlbrace]{align}
    \left. \left< \cfrac{\partial L_{pe}}{\partial \hat{\bm{u}}_{oc}}, \delta \bm{u}_{oc} \right> \middle|_{opt} \right. =
    \left. \left< \cfrac{\partial (F_{pe} + a_{oc} + b_{oc})}{\partial \hat{\bm{u}}_{oc}}, \delta \bm{u}_{oc} \right> \middle|_{opt} \right.= 0, \label{eq:dLpe_duoc}\\
    \left. \left< \cfrac{\partial L_{pe}}{\partial \bm{v}_{ocpe}}, \delta \bm{v}_{ocpe} \right> \middle|_{opt} \right.  =
    \left. \left< \cfrac{\partial (F_{pe} +  a_{oc})}{\partial \bm{v}_{ocpe}}, \delta \bm{v}_{ocpe} \right> \middle|_{opt} \right.= 0, \label{eq:dLpe_dvoc}\\
    \left. \left< \cfrac{\partial L_{pe}}{\partial \hat{\varphi}_{oc}}, \delta \varphi_{oc} \right> \middle|_{opt} \right. =
    \left. \left< \cfrac{\partial (F_{pe} + a_{oc}+b_{oc} + \lambda d_{o})}{\partial \hat{\varphi}_{oc}}, \delta \varphi_{oc} \right> \middle|_{opt} \right.= 0, \label{eq:dLpe_dphioc}\\
    \left. \left< \cfrac{\partial L_{pe}}{\partial v_{\varphi_{ocpe}}}, \delta v_{\varphi_{ocpe}} \right> \middle|_{opt} \right. =
    \left. \left< \cfrac{\partial (F_{pe} + b_{oc})}{\partial v_{\varphi_{ocpe}}}, \delta v_{\varphi_{ocpe}} \right> \middle|_{opt} \right.= 0, \label{eq:dLpe_dvphioc}\\
    \left. \cfrac{\partial L_{pe}}{\partial \hat{\omega}_{oci}}\middle|_{opt} \right. = 
    \left. \left< \cfrac{\partial (F_{pe} +   a_{oc}(\omega_{oci}) )}{\partial \hat{\omega}_{oci}}, \delta \omega_{oci} \right> \middle|_{opt} \right.= 0, \label{eq:dLpe_domegaoc}\\
    \left. \left< \cfrac{\partial L_{pe}}{\partial \hat{\bm{u}}_{sc}}, \delta \bm{u}_{sc} \right> \middle|_{opt} \right. = 
    \left. \left< \cfrac{\partial (F_{pe} + a_{sc})}{\partial \hat{\bm{u}}_{sc}}, \delta \bm{u}_{sc} \right> \middle|_{opt} \right. =0, \label{eq:dLpe_dusc}\\
    \left. \left< \cfrac{\partial L_{pe}}{\partial \bm{v}_{scpe}}, \delta \bm{v}_{scpe} \right> \middle|_{opt} \right. =  
    \left. \left< \cfrac{\partial (F_{pe} + a_{sc})}{\partial \bm{v}_{scpe}}, \delta \bm{v}_{scpe} \right> \middle|_{opt} \right. =0, \label{eq:dLpe_dvsc}\\
    \left. \cfrac{\partial L_{pe}}{\partial \hat{\omega}_{sci}}\middle|_{opt} \right. =  
    \left. \left< \cfrac{\partial (F_{pe} + a_{sc}(\omega_{sci}) )}{\partial \hat{\omega}_{sci}}, \delta \omega_{sci} \right> \middle|_{opt} \right.=0, \label{eq:dLpe_domegasc}
\end{empheq}
\begin{empheq}[left=\empheqlbrace]{align}
    \left. \left< \cfrac{\partial L_{sb}}{\partial \hat{\bm{u}}_{oc}}, \delta \bm{u}_{oc} \right> \middle|_{opt} \right. =
    \left. \left< \cfrac{\partial (F_{sb} + a_{oc} + b_{oc})}{\partial \hat{\bm{u}}_{oc}}, \delta \bm{u}_{oc} \right> \middle|_{opt} \right.= 0, \label{eq:dLsb_duoc}\\
    \left. \left< \cfrac{\partial L_{sb}}{\partial \bm{v}_{ocsb}}, \delta \bm{v}_{ocsb} \right> \middle|_{opt} \right.  =
    \left. \left< \cfrac{\partial (F_{sb} +  a_{oc})}{\partial \bm{v}_{ocsb}}, \delta \bm{v}_{ocsb} \right> \middle|_{opt} \right.= 0, \label{eq:dLsb_dvoc}\\
    \left. \left< \cfrac{\partial L_{sb}}{\partial \hat{\varphi}_{oc}}, \delta \varphi_{oc} \right> \middle|_{opt} \right. =
    \left. \left< \cfrac{\partial (F_{sb} + a_{oc}+b_{oc} + \lambda d_{o})}{\partial \hat{\varphi}_{oc}}, \delta \varphi_{oc} \right> \middle|_{opt} \right.= 0, \label{eq:dLsb_dphioc}\\
    \left. \left< \cfrac{\partial L_{sb}}{\partial v_{\varphi_{ocsb}}}, \delta v_{\varphi_{ocsb}} \right> \middle|_{opt} \right. =
    \left. \left< \cfrac{\partial (F_{sb} + b_{oc})}{\partial v_{\varphi{ocsb}}}, \delta v_{\varphi{ocsb}} \right> \middle|_{opt} \right.= 0, \label{eq:dLsb_dvphioc}\\
    \left. \cfrac{\partial L_{sb}}{\partial \hat{\omega}_{oci}}\middle|_{opt} \right. = 
    \left. \left< \cfrac{\partial (F_{sb} +   a_{oc}(\omega_{oci}) )}{\partial \hat{\omega}_{oci}}, \delta \omega_{oci} \right> \middle|_{opt} \right.= 0, \label{eq:dLsb_domegaoc}\\
    \left. \left< \cfrac{\partial L_{sb}}{\partial \hat{\bm{u}}_{sc}}, \delta \bm{u}_{sc} \right> \middle|_{opt} \right. = 
    \left. \left< \cfrac{\partial (F_{sb} + a_{sc})}{\partial \hat{\bm{u}}_{sc}}, \delta \bm{u}_{sc} \right> \middle|_{opt} \right. =0, \label{eq:dLsb_dusc}\\
    \left. \left< \cfrac{\partial L_{sb}}{\partial \bm{v}_{scsb}}, \delta \bm{v}_{scsb} \right> \middle|_{opt} \right. =  
    \left. \left< \cfrac{\partial (F_{sb} + a_{sc})}{\partial \bm{v}_{scsb}}, \delta \bm{v}_{scsb} \right> \middle|_{opt} \right. =0, \label{eq:dLsb_dvsc}\\
    \left. \cfrac{\partial L_{sb}}{\partial \hat{\omega}_{sci}}\middle|_{opt} \right. =  
    \left. \left< \cfrac{\partial (F_{sb} + a_{sc}(\omega_{sci}) )}{\partial \hat{\omega}_{sci}}, \delta \omega_{sci} \right> \middle|_{opt} \right.=0, \label{eq:dLsb_domegasc}
\end{empheq}
where the expressions within the brackets represent the directional derivatives of the functional.
The variables denoted with a hat, such as $\hat{\bm{u}}$, represent the variables with respect to which the variational analysis is performed and should not be confused with the state variables, which are the solutions of the state equations. This distinction is made to clearly differentiate between the variables used in the sensitivity analysis and the state variables.

Eqs. (\ref{eq:dLpe_duoc}), (\ref{eq:dLpe_dphioc}), and (\ref{eq:dLpe_domegaoc}) 
can be used to derive $\bm{v}_{ocpe}$, $\bm{v}_{ocsb}$, $\bm{v}_{scpe}$, $\bm{v}_{scsb}$, $v_{\varphi{ocpe}}$, and $v_{\varphi{ocsb}}$ as follows:
\begin{align}
    \bm{v}_{ocpe} &= c_{ocpe}\bm{u}_{oc},  \quad \bm{v}_{scpe} = c_{scpe}\bm{u}_{sc}, \quad v_{\varphi_{ocpe}} = c_{ocpe}\varphi_{oc}, \\
    c_{ocpe} &=  \frac{-\alpha_{pe}\omega_{sc}^2}{(\omega_{oc}^2 - \omega_{sc}^2)^2} + (1 - \alpha_{pe})\sum^{n}_{i=1}\frac{(\omega_{oci} - \bar{\omega}_i)}{\omega_{oci}\bar{\omega}_i^2}, \\
    c_{scpe} &=   \frac{\alpha_{pe}\omega_{oc}^2}{(\omega_{oc}^2 - \omega_{sc}^2)^2}.
\end{align}
\begin{align}
    \bm{v}_{ocsb} &= c_{ocsb}\bm{u}_{oc},  \quad \bm{v}_{scsb} = c_{scsb}\bm{u}_{sc}, \quad v_{\varphi_{ocsb}} = c_{ocsb}\varphi_{oc}, \\
    c_{ocsb} &=  \frac{-\alpha_{sb}\omega_{sc}^2}{(\omega_{oc}^2 - \omega_{sc}^2)^2} + (1 - \alpha_{sb})\sum^{n}_{i=1}\frac{(\omega_{oci} - \bar{\omega}_i)}{\omega_{oci}\bar{\omega}_i^2} ,\\
    c_{scsb} &=   \frac{\alpha_{sb}\omega_{oc}^2}{(\omega_{oc}^2 - \omega_{sc}^2)^2}.
\end{align}

From the above equations, the sensitivities for updating ${\phi_{p \psi}}$ and ${\phi_{s \psi}}$ can be obtained as follows:
\begin{align}
    \begin{split}
        {F}'_{pe} =\frac{\partial L_{pe}}{\partial {\phi_{p \psi}}} = &c_{ocpe} \left\lbrace \frac{\partial a_{oc}({\phi_{p \psi}}, {\phi_{s \psi}}, \omega_{oc}, \bm{u}_{oc}, \bm{u}_{oc}, \varphi_{oc})}{\partial {\phi_{p \psi}}} 
        + \frac{\partial b_{oc} ({\phi_{p \psi}}, \bm{u}_{oc}, \varphi_{oc}, \varphi_{oc})}{\partial {\phi_{p \psi}}} \right\rbrace \\
        &+ c_{scpe} \left\lbrace \frac{\partial a_{sc}({\phi_{p \psi}}, {\phi_{s \psi}}, \omega_{sc}, \bm{u}_{sc}, \bm{u}_{sc})}{\partial {\phi_{p \psi}}}  \right\rbrace + \lambda ,\\
        =&c_{ocpe} \left\lbrace  a_{oc}(1, {\phi_{s \psi}}, \omega_{oc}, \bm{u}_{oc}, \bm{u}_{oc}, \varphi_{oc})
        +  b_{oc} (1, \bm{u}_{oc}, \varphi_{oc}, \varphi_{oc}) \right\rbrace \\
        &+ c_{scpe} a_{sc}(1, {\phi_{s \psi}}, \omega_{sc}, \bm{u}_{sc}, \bm{u}_{sc}) + \lambda,
    \end{split}\\
    \begin{split}
        {F}'_{sb} = \frac{\partial L_{sb}}{\partial {\phi_{s \psi}}} = &c_{ocsb} \left\lbrace \frac{\partial a_{oc}({\phi_{p \psi}}, {\phi_{s \psi}}, \omega_{oc}, \bm{u}_{oc}, \bm{u}_{oc}, \varphi_{oc})}{\partial {\phi_{s \psi}}} 
        + \frac{\partial b_{oc} ({\phi_{p \psi}}, \bm{u}_{oc}, \varphi_{oc}, \varphi_{oc})}{\partial {\phi_{s \psi}}} \right\rbrace\\
        &+ c_{scsb} \left\lbrace \frac{\partial a_{sc}({\phi_{p \psi}}, {\phi_{s \psi}}, \omega_{sc}, \bm{u}_{sc}, \bm{u}_{sc})}{\partial {\phi_{s \psi}}}  \right\rbrace,\\
        =&c_{ocsb} \left\lbrace  a_{oc}({\phi_{p \psi}}, 1, \omega_{oc}, \bm{u}_{oc}, \bm{u}_{oc}, \varphi_{oc}) 
        \right\rbrace \\
        &+ c_{scsb} \left\lbrace a_{sc}({\phi_{p \psi}}, 1, \omega_{sc}, \bm{u}_{sc}, \bm{u}_{sc})  \right\rbrace .
    \end{split}
\end{align}

%% The Appendices part is started with the command \appendix;
%% appendix sections are then done as normal sections

%% If you have bibdatabase file and want bibtex to generate the
%% bibitems, please use
%%
\bibliographystyle{elsarticle-num-names} 
\bibliography{mybibfile}

%% else use the following coding to input the bibitems directly in the
%% TeX file.

% \begin{thebibliography}{00}

% %% \bibitem[Author(year)]{label}
% %% Text of bibliographic item

% \bibitem[ ()]{}

% \end{thebibliography}
\end{document}